 \documentstyle[epsfig]{jfm}

\def\bigcirc{open \; circles}
\def\bullet{closed \; circles}
\def\triangle{open \; triangles}
\def\blacktriangle{closed \; triangles}
\def\square{open \; squares}
\def\blacksquare{closed \; squares}
\def\boxplus{crossed \; boxes}
\def\Diamond{diamonds}
\def\times{crosses}

\begin{document}
\title[Local measurement of bubble deformation
]{Foam in a two-dimensional Couette shear: \\
a local measurement of bubble deformation. }

\author[E. Janiaud and F. Graner]{
E. Janiaud$^{1,2}$ \and F. Graner$^{3}$
\thanks{$\;$to whom correspondence
should be addressed at graner@ujf-grenoble.fr}
}

\affiliation{$^1$
   Laboratoire des Milieux D\'esordonn\'es et
     H\'et\'erog\`enes, case 78, Universit\'e Paris
6 and CNRS UMR 7603, 140 rue de Lourmel, 75015 Paris, 
France.\\[\affilskip]
$^2$
  Universit\'e Paris 7, Denis Diderot, F\'ed\'eration de Recherche
FR2438 Mati\`ere et Syst\`emes Complexes, 2 place Jussieu, 75251 Paris Cedex
05, France.
\\[\affilskip]
$^3$
Laboratoire de Spectrom\'etrie Physique,
CNRS UMR 5588 et Universit\'e  Grenoble I,
BP 87, F-38402 St Martin d'H\`eres Cedex, France.}


\maketitle


\begin{abstract}
We re-analyse experiments on a foam sheared in a two-dimensional 
Couette geometry [Debr\'egeas {\it et al.}, Phys. Rev. Lett. {\bf 87},
178305 (2001)]. We characterise the bubble deformation by a texture
tensor. Our measurements are local in time: they show two regimes,
one transient and one stationary. They provide both the
average and fluctuations of the anisotropy.
Measurements are also local in space: they show that both the 
deformation and the elastic contribution to the stress field do
not localise, varying smoothly across the shear gap. We can thus 
describe the foam
as a continuous medium with elastic properties.
\end{abstract}

\vskip 2\baselineskip

  \begin{figure}
     \centering
\includegraphics[width=0.9\textwidth]{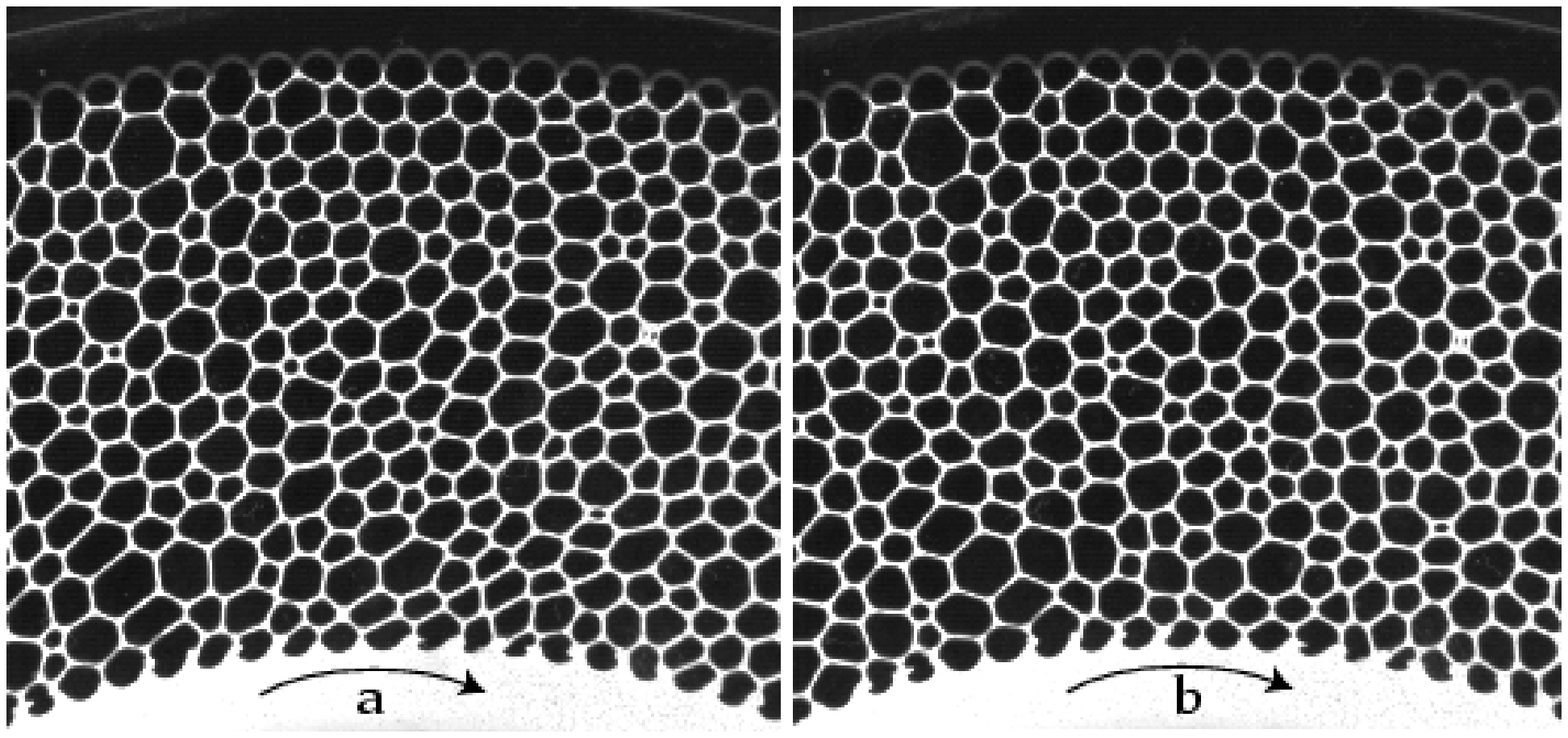}
\includegraphics[width=0.9\textwidth]{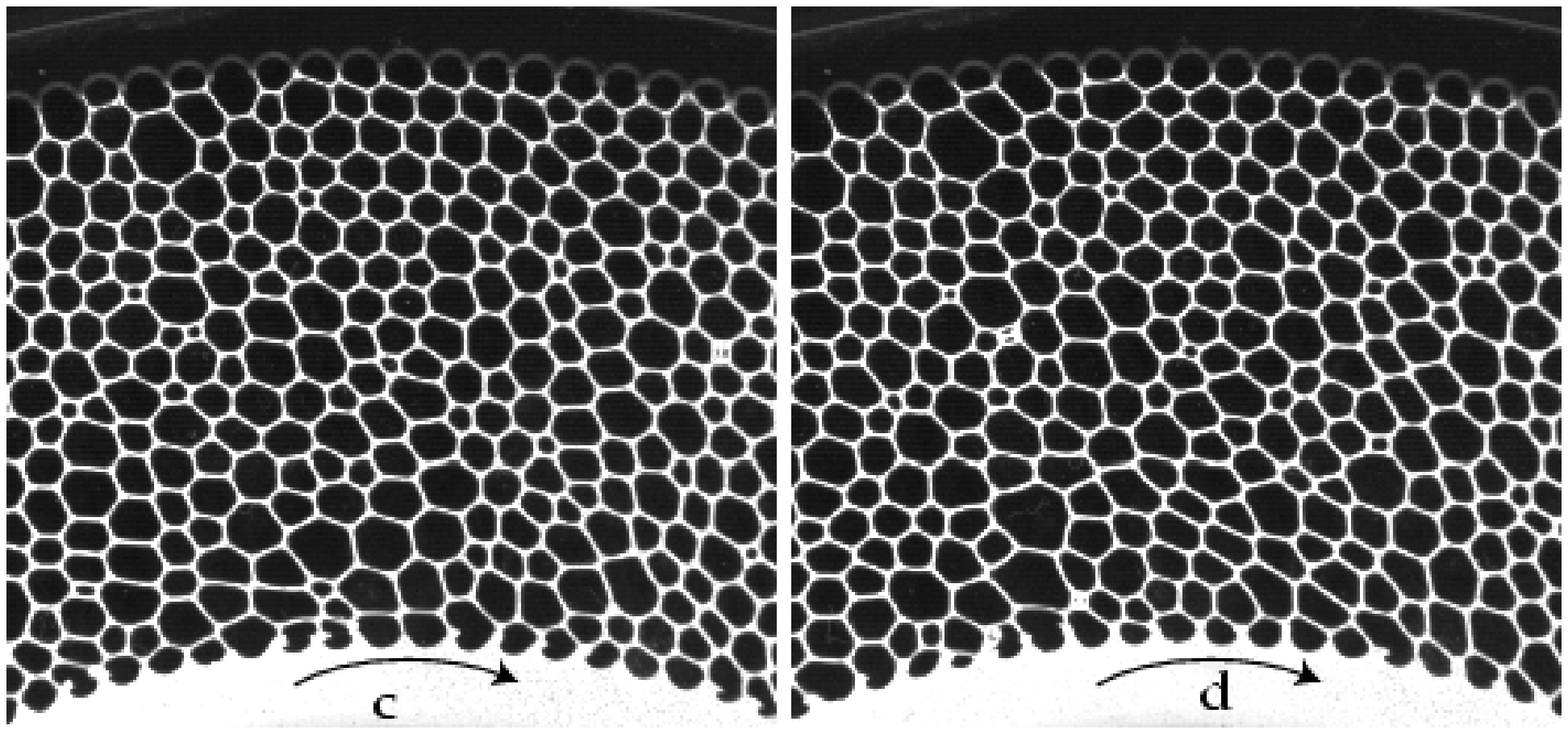}
\caption{Run 1 (transient regime): foam under steady quasistatic shear in a
Couette geometry. Pictures of the foam correspond to different values of applied shear strain $\gamma$:
(a) 0, immediately after switching from counter-clockwise to clockwise rotation; (b)  0.15; (c) 
0.5, when the inner  boxes have reached the yield strain
(see Fig.  \protect\ref{Gamma s});  and (d) 7.6, well into the stationary regime.
The cross-component ${r\theta}$ of the texture tensor, defined in eq. (\protect\ref{defM}), is  positive (a), zero (b), negative (c), and  stationary (d).
  }
\label{images}
  \vspace{1cm}
\end{figure}

  \section{Introduction}

A liquid foam exhibits   ``complex" behaviour under stress:  it is
elastic for small deformation, plastic for large deformation, and flows
at large deformation rates \cite{hutzler,jiangpre}. It is a
model for the study of
complex
fluids: its typical length scale permits direct
observation. In particular, a foam with only one bubble layer 
(so-called ``two-dimensional foam'' \cite{hutzler})  is easy to image,
and image analysis yields information on all the  geometrical
properties of the foam.

Recently, Debr\'egeas {\it et al.} \cite{debregeas1} 
quasistatically sheared a fairly dry  foam in a two-dimensional  Couette
geometry (Fig. \ref{images}).
They studied the velocity field and showed that it  localises
in a shear band:
its changes of value happen as a steep gradient 
on a layer much thinner than the gap, rather than over a long distance. 
It decreases exponentially across the gap $(r_1-r_0)$, as $\exp\left[-(r-r_0)/d\right]$. Here $r$ is the
distance to the wheel center, $r_0$ the radius of the  inner wheel, $r_1$ the radius of the outer wheel,
and $d \ll (r_1-r_0)$ is a characteristic length
comparable to one bubble diameter. Such
shear-banding
has been explained and numerically simulated by the combination of
frozen disorder,   elastic
behaviour at low deformation,  and local discrete bubble 
rearrangements, leading to an instability  more likely to occur at 
the inner side where fluctuations are larger
\cite{debregeas2}.

What dominates the macroscopic behaviour of the material: averages or 
fluctuations?
This question  is important in the rheology of foams, as well as in 
other disordered materials.
Briefly, if the signature of fluctuations averages out at large 
scales, a ``thermodynamical limit" exists in which the average 
dominates; a constitutive equation relating 
the stress and strain fields can then describe the material  as a continuous medium 
(like a solid or a liquid).
If the fluctuations have the same order of magnitude as the average 
at all scales, they can dominate the physical behaviour, which need 
not correspond to any continuous medium.
Whenever the fluctuations dominate, we can ask  
 whether they   correlate  (leading to large stress drops) or have a white spectrum.
The answer need not be the same for different materials (say, foam 
and granular materials)
nor for different quantities (say, velocity and deformation).

Similarly, the existence and origin of localisation raise different 
questions: the effect of boundary conditions, 
the respective roles of strain and strain rate, the precise role of fluctuations, the differences between 
various (disordered) materials, the distinction between sharp \cite{coussot} or exponential
\cite{debregeas1} decreases,
 the dimension of 
space (2D or 3D), and even the field itself (velocity or 
deformation). 
We would like to extract additional  physical information, beyond the velocity field.

Most suggestions to quantify  statistically   the anisotropy  and the
deformation
of the internal structure of the foam, that is, the
bubble wall network, are based on scalar quantities
\cite{elias}.
Since the deformation encompasses information on both anisotropy and
orientation,
it is actually a tensor  \cite{landau,alexander}.
Recent suggestions of possible tensorial descriptions
based on bubble details are promising
\cite{kraynik2000,kraynik2003,ball,blumenfeld}.
Here, we use
the ``texture tensor'' (eq. \ref{defM}), which has  the following
advantages
\cite{aubouy}: (i) it is purely geometrical, independent of stresses
and forces; (ii) it applies
to both small and  large numbers of bubbles; (iii) it is local in time
and space \cite{delft}, so it can
characterise a sub-region of the foam; (iv)  it applies to disordered 
or ordered, 2D or 3D materials, in elastic, plastic or fluid regimes.

The plan of this paper is as follows.
Section (\ref{MatMet}) reviews the original  experimental set-up, and
our image analysis protocol. The theoretical section 
(\ref{theory}) reviews the definitions and measurements
of stress, texture and strain tensors; it can be read separately.
The next section (\ref{results})  presents our analyses of
experiments.  
The first (section \ref{transient}), 
resolved in both time and space, studies the transient regime before a
stationary flow is
established. The second (section \ref{permanent}), resolved in space but  averaged in time,
shows that in 
stationary flow, both the bubble anisotropy and the elastic contribution to the
deviatoric stress field vary smoothly across the
Couette gap; they do not
localise near the inner rotating wheel. The third (section \ref{fluctuat}),
 again resolved in time,
shows  that in stationary flow the fluctuations are larger near the
inner wheel.  In section (\ref{discu}), we discuss whether we can describe the foam
as a continuous medium with elastic properties.  

  \section{Materials and Methods} \label{MatMet}
  \subsection{ Experiments}

Georges Debr\'egeas generously provided experimental data  obtained  in two
different runs. For  experimental   details see  Ref. 
\cite{debregeas1}.
Briefly,  the set-up  consists of  an inner shearing wheel and an outer fixed one, with no-slip conditions due to 
tooth-shaped boundaries (tooth depth: radius 1.2 mm).
The foam is confined
between
two transparent plates separated by $h=2$~mm spacers (Fig. \ref{FigPhoto}(d)); this foam 
thickness is smaller than
the
distance between bubble centers (see below),
enforcing a single layer of bubbles 
\cite{Cox}.
The fluid fraction, measured by the weight of water
introduced in the cell, is $\Phi=5.2\%$. The bubble size dispersity 
is $\sim 30\%$,
enough to prevent crystallisation.
Coarsening and size sorting under shear are negligible during the
experiment. 
A CCD
digital camera records the position and the shape of the bubbles.

The inner wheel rotates  slowly enough ($V_{wheel}=0.25$~mm.s$^{-1}$)
that the flow is well in quasistatic regime: \label{qstat}
the foam passes through a succession of equilibrium states.
All averaged static physical quantities, like the  elastic  strain or 
stress,  remain the same if the applied
shear rate 
$V_{wheel}/(r_1-r_0)$
varies.
Kinetic quantities, like the 
velocity field, the rate of T1s,  and the input power and 
dissipation, scale as the shear rate.
This scaling is typical of  solid-like friction (only at 
higher velocity does the foam exhibit  a fluid-like friction, when 
the input power and dissipation scale like the square of the 
shear rate).
As
required for comparison to other experiments,
we thus express all results  as a
function  of the total applied shear strain: 
\begin{equation}
\gamma= \frac{V_{wheel}\; t}{r_1-r_0}.
\label{prop_time}
\end{equation}
Note that $\gamma$ is 
 proportional to time $t$, 
not to the local shear in each box.

Run 1 (unpublished data)  contains a transient
regime. To prepare the foam, the inner disc is rotated counter-clockwise, 
until
a stationary regime is reached. Then, at an arbitrary time,  chosen as the origin
($t=\gamma=0$), the shear direction is switched 
to clockwise (Fig. \ref{images}),
the experiment
begins and  500 pictures are recorded.
The $r\theta$ components
of the deformation thus switch from positive to negative (Fig. \ref{FigTransi}).
Each picture shows 200 bubbles, in an angular 
sector of the experimental cell (Fig. \ref{images}).
The average distance between the centres of mass of  neighbouring bubbles
is $3.0$ mm, with a standard deviation of $\pm 0.5$ mm, corresponding to a
mean bubble wall length of $\langle \ell \rangle = 1.7$ mm.
The internal 
radius is 
$r_0=71$~mm (same for both runs) and the external radius is $r_1=112$~mm
(different for each run).
Thus $\gamma=1$ corresponds
to a wheel displacement of $ V_{wheel}\; t = r_1-r_0=41$~mm: {\it 
i.e.} 15-20 bubble
diameters. 

\begin{figure}
\includegraphics[width=0.92\textwidth]{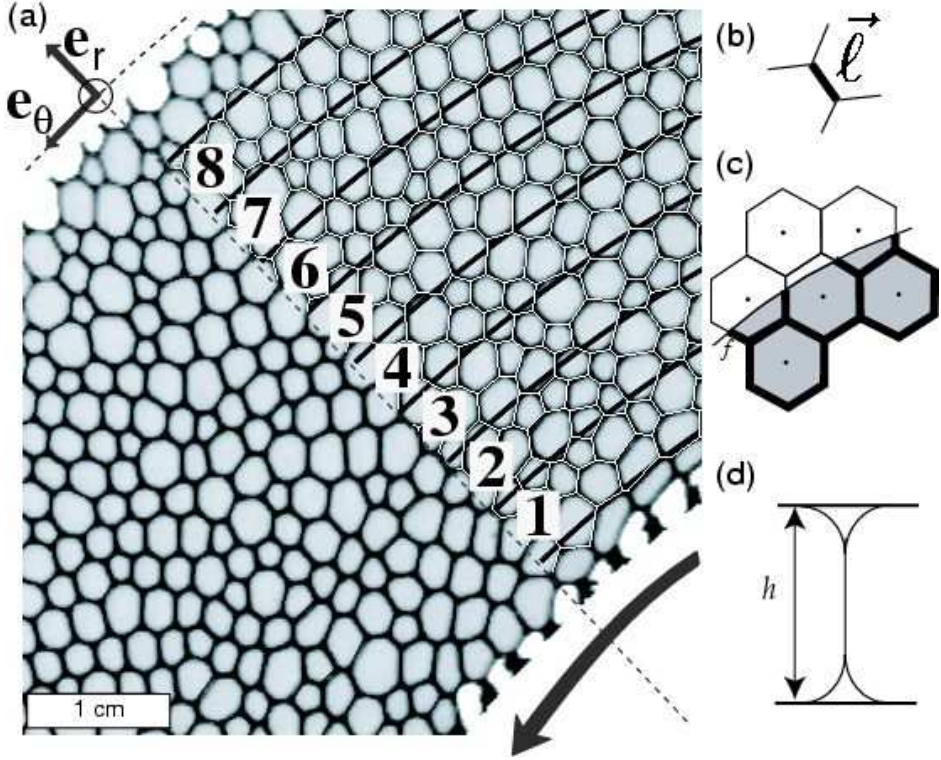}
\caption{ Run 2 (stationary flow): image analysis and notations. (a)  {\it Bottom left:} top view of the original
experiment by Debr\'egeas {\it
et al.} \protect\cite{debregeas1}, rotating counter-clockwise. 
    {\it Top right:} image analysis. The reconstructed network of
vertices linked by vectors
(white) and the boundaries of the 8 equi-spaced concentric circular boxes used 
as representative volume elements (RVE) for spatial
analysis (black circular arcs) are superimposed on the snapshot of the experiment. We exclude the regions near both
wheels, where the bubbles are not entirely visible.
(b) We denote by $\vec \ell$ the vector linking both vertices of a given wall.
  (c) If one of these vertices is outside of the box (defined
in (a), top right), we denote by $f \in [0,1]$ the
fraction   of the bubble wall length inside the box; $f=0$ if the wall is 
entirely outside, $f=1$ if it is entirely inside. (d) Side-view sketch of a vertical bubble wall cross-section, between two horizontal glass plates.
}
\label{FigPhoto}
\end{figure}

Run 2 (published  in
Ref. \cite{debregeas1})  focusses on the  stationary regime. 
The preparatory rotation is clockwise. 
Then, at an arbitrary time,  chosen as the origin
($t=\gamma=0$), it  is switched to counter-clockwise
 (Fig. \ref{FigPhoto}).
Pictures  are recorded only after 
a full $2\pi$-turn, corresponding to
$\gamma= 8$; the
  $r\theta$ components
of the deformation  are thus   positive (Fig. \ref{FigPerm}).
A quarter of the Couette cell is filmed, with 700 bubbles in
each picture.
To improve the statistics, the camera is then displaced to record successively
the four quarters;  2000 pictures are obtained as four 500-images movies.
Data are measured  and  averaged over the four 
quarters of the foam.
The average distance
between the centres of mass of  neighbouring bubbles is $2.4$ mm, with 
a standard
deviation of $\pm 0.4$ mm,  corresponding to a mean bubble wall length of
$\langle \ell \rangle = 1.4$ mm. 
The internal 
radius is 
$r_0=71$~mm (same for both runs) and the external radius is $r_1=122$~mm
(different for each run).
Thus $\gamma=1$ corresponds
to a wheel displacement of $51$~mm: {\it i.e.} 20-25 bubbles
diameters.


  \subsection{Image analysis}
We analyse the images using a home-made extension module to  the
public software
NIH-Image.  We threshold the image, and skeletise the
bubble walls to a one-pixel
thickness. We identify the points where three walls meet (``vertices'').
We replace each wall by the vector $\vec \ell$ linking its vertices
(white lines in Fig. \ref{FigPhoto}a, top right).

To respect the symmetry of
the Couette cell,
this  wall vector $\vec \ell$ (Fig. \ref{FigPhoto}b) is expressed in local
polar coordinates  $(r,\theta)$
 (Fig. \ref{FigPhoto}a). This neglects the variation of the polar 
referential between both its end points. This assumption is correct as long as   $\ell \ll r$.

For the same reason, we choose equi-spaced orthoradial, circular boxes (Fig.
\ref{FigPhoto}a, top right) to divide the foam and  measure
the deformation using spatial averages  to improve the statistics.
In the quasistatic regime, the bubble deformation remains very 
small and barely visible by eye. After trying  various box sizes, we choose  8 boxes as a compromise 
between the quality of the statistics and the level of visible 
details.

For Run 1, each box is $ 2.7 \langle \ell\rangle $ wide,
 and box number $i$ is at position $r_1^{i}=r_0 + [0.4+2.7
(i-0.5)] \langle \ell \rangle $.
The number of bubbles in each box grows linearly, 
typically from 17 bubbles in the 1st box, to 32 in the 8th box. 

For Run 2, $r_2^{i}=r_0 + [2.8+3.1(i-0.5)] \langle\ell \rangle $. 
Due to the shape of the pictures (Fig. 
\ref{FigPhoto}a), the number of bubbles first increases with $i$ (from 
typically 105 bubbles in the 1st box), then decreases (to 41 in the 
8th box).

\section{Definitions and measurements of deviatoric elastic stress and strain} \label{theory}

  \subsection{Capillary contribution to the stress} \label{presentation stress}

\subsubsection{Theoretical definition}

In fluid foams, the 
three main contributions to stress $\overline{\overline{\sigma}}$ 
are the capillary term 
$\overline{\overline{\sigma}}^{cap}$
due to the tension of the liquid  walls, the 
pressure of the gas within each bubble, and the viscous dissipation%
.
In what follows,  the ``deviatoric" part of the capillary stress refers to the traceless tensor, {\it i.e.}
$\overline{\overline{\sigma}}^{cap}- {\rm Tr}\left(\overline{\overline{\sigma}}^{cap}\right)
 \overline{\overline{I}}/2$, where $ \overline{\overline{I}}$ is 
the identity tensor  in 2 dimensions. It consists in the  
shear stress $\sigma^{cap}_{r\theta}$ and the normal stress difference 
$\sigma^{cap}_{rr}-\sigma^{cap}_{\theta\theta}$.

In the extreme ``wet foam" limit $\Phi  \sim 1$, bubbles are well separated from each other. The capillary contribution
in a representative volume element (RVE) ${{V}}$ which contains 
$N$  bubbles with 
surface tension $\Gamma$ is
\cite{rosenkilde,batchelor}:
\begin{eqnarray}
\overline{\overline{\sigma}}^{cap} &=& \frac{\Gamma}{{{V}}} \; 
\sum_{S \in {{V}}} \oint_S \left(\overline{\overline{I}} - \vec{n} 
\otimes \vec{n} \right) \; ds,
\nonumber\\
\sigma_{ij}^{cap} &=&  \Gamma \; \frac{N}{{{V}}} \; \left\langle \oint_S 
\left(\delta_{ij} - n_in_j \right) \; ds
\right\rangle_{{V}}.
\end{eqnarray}
Here  $ N/{{V}}$ is the density of bubbles; $S$ is the surface of each bubble; 
$\vec{n}$ is the unit vector normal to the surface element $ds$;  and 
$\langle . \rangle_{{V}}$ is the average over bubbles in ${{V}}$.

In the other extreme, ``dry foam" limit $\Phi  \ll 1$, pairs of  bubble surfaces merge into a 
film with surface tension $2 \Gamma$. The capillary contribution is
\cite{khan1986,khan1987,kraynik2000,kraynik2003}:
\begin{equation}
\overline{\overline{\sigma}}^{cap} = \frac{2 \Gamma}{N{{V}}} 
\oint_S  \left(\overline{\overline{I}} - \vec{n} \otimes \vec{n}\right) ds
+ {\cal O}\left( \Phi^2\right).
\label{2gamma}
\end{equation}
The correction of order $\Phi^2$ is a contribution of the integral over 
the surface $S$ of its mean 
curvature \cite{mecke}. Since the mean curvature is mostly concentrated in bubble edges
(``Plateau borders"), this contribution  is physically equivalent to the recently identified  line tension 
which affects the bubble edges \cite{kern,geminard}.

\subsubsection{Operational definition in two dimensions} \label{operational}

A true 2D foam has an interesting property: at equilibrium, or in a 
quasistatic regime, the coarse-grained Laplacian of the pressure is zero 
\cite{graner2001,wuppertal}. 
The small scale fluctuations of   bubble pressures,  according to bubble side numbers,
 average out to zero at large scales \cite{hutzler}.
In the present Couette geometry,  where no  large-scale pressure 
gradient exists,  pressure does not vary significantly
\cite{morgan}.
Hence,  in-plane  curvatures remain much smaller than the inverse of bubble 
sizes. They contribute to the stress through their algebric
average (not their second moment), which is small along
any path within the foam:  curvatures are counted positively if the path crosses the wall from the concave side \cite{wuppertal}.
They thus induce only a negligible correction to the deviatoric 
capillary stress, with respect to
straight walls
\cite{asipauskas}.

However,  adapting  the above definition (eq. \ref{2gamma}) to 
measure deviatoric capillary stress fields in  bubble monolayers requires care. In 
the present case, the curvature radius in the third dimension, not 
visible in images, is smaller or equal to the half distance between plates
(Fig. \ref{FigPhoto}(d)).
Its inverse,  namely the curvature in the third dimension, is larger
than the curvature visible in the image.  This case is 
thus intermediate between two limiting  bubble monolayers: bubble 
rafts, and Langmuir foams.

In bubble rafts, stress has already been measured, in a 
circular Couette geometry,  by mechanical measurements of the force 
at the boundaries \cite{pratt}, or indirectly by local measurements of bubble 
ellipticity \cite{ybert}. Since  bubble  top surfaces are curved (they are 
approximately spherical caps), direct local measurements of stress are 
difficult.

At the other extreme, Langmuir foams are one molecule thick: their 
aspect ratio is of the order of nanometers divided by centimeters, with
 no well-defined curvature in the third dimension. They are 
thus probably the closest approximation to a true 2D foam (as long as 
3D viscosity effects remain negligible), and capillary stress fields 
can be measured as described below (eq. \ref{sigma cap})
\cite{langmuir_foam}.

In the present case, 
for a single  layer  of bubbles   \cite{Cox}  in the quasistatic regime,
the dry part of the film remains vertical  (Fig. \ref{FigPhoto}(d)).
By symmetry, 
we can assume that
  the curvatures on both sides of a film are almost opposite, so that 
their effects almost cancel each other.  
We thus neglect the effect of curvature in 
the third dimension.  

We simplify   further
eq. (\ref{2gamma}) using two approximations.  First, although strictly speaking
it  applies
 to closed surfaces
(complete surfaces of bubbles), we extend it to averages on open surfaces (walls between bubbles),
assuming that any possible non-compensated  term (due for instance to bubbles at the foam boundary) has a negligible contribution to the average \cite{kraynik2003}.
Second, on closed surfaces 
in 2D, the integral of the mean curvature only contributes a 
constant, so that the ${\cal O}\left( \Phi^2\right)$ term 
need not be considered here;  in what follows, on open surfaces, we again neglect 
the contribution of possible non-compensated  terms.

Finally, eq. (\ref{2gamma}) 
simplifies  to an expression which has already been used (but not always justified) for 
various types of bubble monolayers 
\cite{hutzler,debregeas2,asipauskas,langmuir_foam}:
\begin{equation}
\overline{\overline{\sigma }}^{cap} =\frac{1}{{{V}}} \sum_{\vec{\ell}}
f\; \left( \lambda  \hat{e} \otimes \vec{\ell}\right).
\label{sigma cap dim}
\end{equation}
Here  (Fig. \ref{FigPhoto}) the 2D RVE   ${{V}}$ 
is a surface area; the $\vec{\ell}$s are the chords linking  two 
connected vertices, approximating bubble walls;  
$f$ is the fraction of the wall length contained in ${{V}}$; 
$\hat{e}= \vec{\ell}/\ell$ is the unit vector tangent to $\vec{\ell}$.
The line tension $\lambda$  is equal to $\Gamma h$ multiplied by a geometric
constant which depends on $\Phi$  (Fig. \ref{FigPhoto}(d)): it varies from 2 for a dry foam (two flat interfaces of length $h$)
to $\pi$ for a wet foam (two half-circles of diameter $h$).

\subsubsection{Comments on this definition} \label{comments stress}

Eq. (\ref{sigma cap dim})   is analogous 
to the expression
for the elastic stress in a (2D or 3D) network with a density $\rho$ 
of links $\vec{\ell}$ undergoing  two-body  interactions with 
tensions $\vec{\tau}$ \cite{kruyt}:
\begin{equation}
\overline{\overline{\sigma }}^{cap}= \rho \left\langle
\vec{\tau} \otimes \vec{\ell} \right\rangle_{{V}}.
\label{sigma gene}
\end{equation}

In the continuum limit, the average $\langle . \rangle_{{V}}$ is the average of the wall probability distribution function, and 
this expression for the stress is exactly the same (see for instance 
ref. \cite{kruyt} or the appendix of ref. \cite{aubouy}) as the 
classical definition based on the links crossing the boundaries of 
${{V}}$ \cite{landau}. This equivalence validates {\it a posteriori} our 
application of eq. (\ref{2gamma}) to open surfaces. In 
practice, experimental data have a finite size:
the  number of links used in the calculation, and so the 
signal/noise ratio of the resulting measurement \cite{asipauskas}, is much 
larger  if we use all links in ${{V}}$, rather than only the 
links crossing  its boundaries.

Eq. (\ref{sigma gene}) shows that, in a network of 
points which interact 
through a pairwise potential, we can  measure
 the stress from an image  if we know the expression for the two-points force law 
$\vec{\tau}= \vec{\tau}(\vec{\ell})$  explicitly 
 (for instance Lennard-Jones, or 
harmonic).

 For quasistatic 2D foams (eq. \ref{sigma cap dim}), all tensions 
have the same modulus $\lambda$; hence 
$\vec{\tau}(\vec{\ell}) =   \lambda \hat{e}$. 
The image 
provides 
$\overline{\overline{\sigma }}^{cap}/\lambda$, which   is enough for 
our purposes:
\begin{eqnarray}
\frac{\overline{\overline{\sigma }}^{cap}}{\lambda}&=&
\rho  \left\langle
\hat{e} \otimes \vec{\ell}
\right\rangle_{{V}}
,\nonumber\\
\frac{\sigma^{cap}_{ij}}{\lambda}&=&
\rho  \left\langle \frac{\ell_{i}\ell_{j}}{\ell}
\right\rangle_{{V}} .
\label{sigma cap}
\end{eqnarray}
Here
$\rho$ is the area density of bubble walls: the number of bubble walls 
in ${{V}}$ weighted by $f$, divided by ${{V}}$;
 $\langle . 
\rangle_{{V}}$ is the average over the walls in  ${{V}}$, also 
weighted by $f$.

\subsection{Status of the elastic strain tensor}\label{presentation deformation}

\subsubsection{Macroscopic definition of the elastic strain}

A macroscopic definition of elastic strain  \cite{macosko,farahani}
is usually a function of the {\it deformation gradient} $\overline{\overline{F}}$:
 \begin{eqnarray}
\overline{\overline{F}}
&=& \left( \frac{\partial \vec{x}}{\partial \vec{x}^0}\right)^t = \left( \vec{\nabla}\; \vec{x} \right), \nonumber \\
F_{ij} &=& \frac{\partial x_i}{\partial x^0_j},
\label{defo_gradient}
\end{eqnarray}
where $\vec{x} $ is the current position  of a point initially at position $\vec{x}^0$.
This deformation gradient $\overline{\overline{F}}$ coincides with the identity $\overline{\overline{I}}$ in a solid translation, and has a determinant 1 in a volume-conserving transformation.

From $\overline{\overline{F}}$, we can define two symmetric tensors $\overline{\overline{\cal U}}$ and $\overline{\overline{\cal V}}$ through a rotation
$\overline{\overline{R}}$ \cite{phan}:
 \begin{equation}
\overline{\overline{F}} = \overline{\overline{R}} \; \overline{\overline{\cal U}} =  \overline{\overline{\cal V}} \; \overline{\overline{R}}.
\label{stretch}
\end{equation}
They are called the {\it right} and {\it left stretch tensors}. They coincide with the identity $\overline{\overline{I}}$ in a solid transformation (translation and rotation), and also have a determinant 1 in a volume-conserving transformation.

In turn, they determine a {\it right} and a  {\it left  deformation tensors}, called the Finger tensor and the
Cauchy-Green tensor, respectively  \cite{phan}:
 \begin{eqnarray}
\overline{\overline{B}}
&=&  \overline{\overline{F}}  \; \overline{\overline{F}}^t = 
\overline{\overline{\cal V}}^2, \nonumber \\
 \overline{\overline{C}}
&=&  \overline{\overline{F}}^t  \; \overline{\overline{F}} = \overline{\overline{\cal 
U}}^2.
\label{defo_tensor}
\end{eqnarray}
They too are symmetric, coincide with the identity $\overline{\overline{I}}$ in a solid transformation (translation and rotation), and have a determinant 1 in a volume-conserving transformation.

Finally, a {\it strain measure tensor}  $\overline{\overline{E}}$ is for instance equal to 
$\left(\overline{\overline{B}}-\overline{\overline{I}}\right)/2$, $\left(\overline{\overline{C}}-\overline{\overline{I}}\right)/2$, or more generally to one of the Seth-Hill strain measure tensors based on the successive powers of 
$ \overline{\overline{\cal U}}$ or $ \overline{\overline{\cal V}}$:
 \begin{eqnarray}
\overline{\overline{E}}^{(m)} &=&
\frac{ \overline{\overline{\cal U}}^m - \overline{\overline{I}}}{m},
\nonumber \\
 {\rm or} \quad 
\overline{\overline{E}}^{(m)} &=&
\frac{ \overline{\overline{\cal V}}^m - \overline{\overline{I}}}{m},
\label{seth_hill}
\end{eqnarray}
where $m$ is a positive or negative integer
\cite{farahani}. They all tend 
towards the linear elasticity definition (the gradient of the displacement field) 
in the limit of infinitesimal  
transformation. 

The particular case of  $m=0$ corresponds to the
  logarithm-based definition \cite{hoger} and generalises the Hencky 
strain  tensor \cite{hencky} to arbitrary deformations:
\begin{eqnarray}
\overline{\overline{E}}^{(0)} &= & 
\log  \overline{\overline{\cal U}} = \frac{1}{2} \; \log \overline{\overline{C}}
, 
\nonumber \\
{\rm or} \quad 
\overline{\overline{E}}^{(0)} &=&  
\log  \overline{\overline{\cal V}} = \frac{1}{2} \; \log \overline{\overline{B}}.
\label{logstrain}
\end{eqnarray}
This {\it true strain tensor} $\overline{\overline{E}}^{(0)} $  correctly quantifies both volume-invariant deformations (which 
appear as traceless strain tensors) and dilations (which appear as 
the logarithm of  the volume, as, for example,  in gases).

Constitutive equations of a material, that is stress-strain relations, can use either of these tensors; some arguments favor right tensors \cite{farahani,larson} while other favor left ones \cite{macosko}.

\subsubsection{Micromechanical definitions of elastic strain}

A microscopic definition of elastic strain 
\cite{kr96,liao,alexander,ball,blumenfeld,kruyt,gold3,zimmerman,aubouy}
adopts a different point of view: it constructs the macroscopic 
elastic strain from
the positions of individual ``sites" ({\it i.e.} atoms, grains, or 
vertices of a network), just as stress is constructed 
from individual forces, see eq. (\ref{sigma gene}).

For instance, Zimmerman considers elastic deformations of 2D or 3D 
ordered atomic crystals
\cite{zimmerman}. He constructs separately the stress (from atomistic 
potentials) and the  elastic strain (from displacements). He shows 
that,  in the ordered cases he considers, his strain yields a 
continuum-mechanics limit  identical to
that derived from  the Cauchy-Green tensor.

Kruyt (2003) expands the analysis he used in a series of papers with 
Rothenburg, which, for instance, predicted the elastic moduli of
disordered or ordered  2D granular materials \cite{kr02}.
He constructs separately the stress from individual forces (obeying 
equilibrium conditions) and the elastic strain from individual 
displacements (obeying geometric compatibility conditions)  small 
enough to avoid rearrangements \cite{kruyt}.

Goldhirsch and Goldenberg  (2002) rewrite elasticity theory on a microscopic 
basis, take the continuous limit, and show its compatibility with 
classical elasticity, including the energetics and conjugation 
relations \cite{gold3}. In particular, they show that their 
definition of elastic strain is better than those based, for instance, 
on fits of the displacements to  a macroscopic strain field 
\cite{liao} and that, even for granular materials, microscopic 
fluctuations   progressively average out when going to macroscopic 
scales.

\subsection{Extension to plastic and fluid regimes}

Macroscopic and micromechanic elastic strain tensors  become difficult to define after the first topological 
rearrangement. They finally lose their meaning after a few rotations of the 
Couette-cell inner wheel, and mathematical singularities appear.
This breakdown of formalism is counterintuitive; experimentally, the mechanical behaviors of elasto-plasto-viscous materials do not appear to change discountinuously. 

We need to treat  the (even large) elastic deformation, independently from the flow  involving
  rearrangements
of the connections between neighbouring sites  (changes in network topology). 
The total applied shear strain  decomposes into plastic and elastic strains.
The plastic strain is much larger, since  it increases linearly 
with time (its derivative, corresponding to the velocity gradient, is 
constant in steady flow). The elastic strain is much smaller:
here, in steady flow, it remains constant; but it is physically 
important, because it characterises the current state of the foam.

We use below the  approach of 
Aubouy {\it et al.} (2003), which consists in three steps. It first characterises the current state of the material by a 
``texture"  (or ``fabric")  tensor  $\overline{\overline{M}}$, built
from statistical averages over microscopic positions, which exists in all regimes. It then defines a reference state characterised by a reference tensor $\overline{\overline{M}}_0$. It finally  
constructs the elastic 
strain by a comparison of both tensors. 
We now review these definitions.
 
\subsubsection{Texture tensor} \label{def_text}

The pattern of the walls in a region of the foam can be statistically 
characterised by the  local  {\it texture tensor} 
$\overline{\overline{M}}$  \cite{aubouy}. In 2D, it writes
(again with averages weighted by $f$):
      \begin{eqnarray}
\overline{\overline{M}}
&=& \left\langle  \vec{\ell}\otimes \vec{\ell} \right\rangle_{{V}}
,\nonumber\\
  M_{ij}&=&
   \langle
 \ell_{i}\ell_{j}
 \rangle_{{V}} 
= \left\langle  \left(
\begin{array}{ll}
\ell_r^2 &\ell_\theta  \ell_r\\
\ell_r\ell_\theta & \ell_\theta^2
\end{array}
\right)\right\rangle_{{V}} .
\label{defM}
\end{eqnarray}
It  accounts for
  the direction and length $\ell$ of the walls;
however, since the symmetry $\vec{\ell} \to -\vec{\ell}$ does not affect the
pattern,
the orientation of $\vec{\ell}$ plays no role.

  The diagonal components  $M_{rr}$ and $M_{\theta\theta}$ of this tensor are
both of
order $\left\langle \ell ^{2}\right\rangle $. Conversely, the
off-diagonal component
$M_{r\theta}=M_{\theta r}$ (the tensor is symmetric), and the
difference   $M_{rr} - M_{\theta\theta}$, together characterise the 
anisotropy: they
are both  much smaller than $\left\langle \ell ^{2}\right\rangle $. They
  vanish
when the foam is isotropic; in which case $\overline{\overline{M}}
= (\langle \ell^2 \rangle /2) \overline{\overline{I}}$ is 
proportional to 
$ \overline{\overline{I}}$, and only 
encodes the r.m.s. wall length.
Hence the tensor $\overline{\overline{M}} $ has two strictly positive
eigenvalues, also of
order $\left\langle \ell ^{2}\right\rangle $, the
larger being in
the direction in which bubbles elongate.
    It thus quantifies the average size, as well as the
anisotropy direction and amplitude, of the network consisting of the 
bubble vertices (Fig. \ref{FigPhoto}a, top right).
In a stationary regime, it is constant, as are all averaged physical quantities.

\subsubsection{Reference state} \label{reference}

At any time,  the current state of the foam  corresponds to   
a current  reference state,  namely 
the  state  that the foam reaches after relaxation  
\cite{porte}.
This reference state defines  
the  {\it reference texture tensor } 
$\overline{\overline{M}}_0$. 
It varies with time; 
but in a stationary regime, it is constant, as are all averaged physical quantities
including $\overline{\overline{M}}$.
The reference state is not necessarily equal to the initial state \cite{porte}.

In the present experiment, letting the inner wheel rotate back freely is difficult,
so that  in practice we do not 
have direct access to a reference state. But we do not need microscopical details
to estimate $\overline{\overline{M}}_0$:
 it suffices to make a physical approximation on the average properties of the foam at rest.
We chose below  to assume  that the reference state  is statistically isotropic.
We thus approximate 
$\overline{\overline{M}}_0$  by
$    M_0 \overline{\overline{I}}$, where 
the constant 
$M_0 = {\rm Tr}  \left(\overline{\overline{M}}_0\right) / 2$ is half the average over all boxes of the squared wall length. 
We implicitly assume that the volume is conserved, which should be 
correct in  Couette flow. More precisely,  to avoid 
overestimating  $M_0$ due to bubble stretching, we take the average 
between $\langle \ell \rangle^2/2$ and $ \langle \ell^2 \rangle/2$;
since here $\langle \ell \rangle=1.40$ mm and $\sqrt{\langle \ell^2 
\rangle}=1.45$ mm, we take $M_0=1.0$ mm$^2$.

\subsubsection{Elastic strain tensor}

We can operationally measure the {\it statistical elastic strain tensor }  $\overline{\overline{U}}$  in each box 
from \cite{aubouy}:
\begin{equation}
\overline{\overline{U}}(\vec r) =
\frac{\log{\overline{\overline{M}}}(\vec r) -
\log{   \overline{\overline{M}}_0       }}{2}.
\label{U log}
\end{equation}

It  obeys \cite{aubouy}
the mathematical requirements for a strain tensor: symmetry 
properties under translation, rotation and index permutation.
It 
is always defined: $\log{\overline{\overline{M}}}$ is the tensor 
with the same eigenvectors as $\overline{\overline{M}}$, but with the 
logarithm of its (strictly positive) eigenvalues.
It is the only dimensionless function of $\overline{\overline{M}}$
and $\overline{\overline{M}}_0$ which is always defined (there is no equivalent of the division for tensors, except when they commute).

It also obeys the physical requirements for an elastic strain tensor. It statistically quantifies 
the deformation reversibly stored in the present state of the 
material. Its differential $d\overline{\overline{U}}$ (and hence the
elastic moduli for infinitesimal deformation around the current state) 
depends explicitly only on the current state.

Moreover, let us consider the case where the (small or large) deformation is elastic 
and affine. Then:
\begin{eqnarray}
\vec{\ell} &=&    \overline{\overline{F}} \; \vec{\ell}_0,\nonumber \\
\vec{\ell} \otimes \vec{\ell} &=& 
\vec{\ell} \;   \vec{\ell}^t =   \overline{\overline{F}} \; \vec{\ell}_0 \;
 \vec{\ell}_0^t \; \overline{\overline{F}}^t 
=  
  \overline{\overline{F}} \; 
\left( \vec{\ell}_0 \otimes \vec{\ell}_0 \right) \; \overline{\overline{F}}^t ,
\nonumber \\
\overline{\overline{M}}&= &\left\langle  \vec{\ell} \otimes  \vec{\ell} \right\rangle
=  \overline{\overline{F}} \; 
\left\langle  \vec{\ell}_0 \otimes \vec{\ell}_0 \right\rangle \; \overline{\overline{F}}^t 
= 
\; \overline{\overline{F}} \; \overline{\overline{M}}_0 \; 
\overline{\overline{F}}^t
.
\end{eqnarray}
If we now assume that $\overline{\overline{M}}_0$ commutes with 
$\overline{\overline{F}}$, which is the case if for instance 
$\overline{\overline{M}}_0$ is isotropic, then
$\overline{\overline{M}}
=   \overline{\overline{M}}_0 \; \overline{\overline{F}} \; 
\overline{\overline{F}}^t$.
We then prove that $\overline{\overline{U}}$
coincides with the true strain  
  \cite{hoger,zimmerman}:
\begin{equation}
\overline{\overline{U}}  =
\frac{\log{\overline{\overline{M}}} -
\log{\overline{\overline{M}}_0} }{2} =
\frac{\log{
\left(
\overline{\overline{M}}_0 \; \overline{\overline{F}} \; 
\overline{\overline{F}}^t
\right)
} -
\log{\overline{\overline{M}}_0} }{2} =
\frac{1}{2}\; 
\log{  \left(\overline{\overline{F}} \;  \overline{\overline{F}}^t \right) }
= \overline{\overline{E}}^{(0)} ,
\end{equation}
see eq. (\ref{logstrain}).
The Finger tensor appears 
because we base our elastic description on a tensor, 
$\vec{\ell} \otimes \vec{\ell} =
\vec{\ell} \; \;  \vec{\ell}^t $. This constrast with 
classical elasticity, based on a scalar product $d\vec{x} \cdot d\vec{x} =
d\vec{x}^t \; d\vec{x} = d\vec{x}_0^t \; \overline{\overline{F}}^t \;  \overline{\overline{F}} \; d\vec{x}_0 $, where the Cauchy-Green tensor appears
\cite{phan}.

A specific feature of the quasistatic regime is that the deformation (and 
hence the elastic strain) remains small, barely visible by eye (Fig. 
\ref{images}), even for large applied shear strain (and hence large 
plastic strain). We thus assume, and check {\it a posteriori} 
(Fig. \ref{U(r)}),  that the components of
$\overline{\overline{U}}$ are much smaller than one, so we approximate 
eq. (\ref{U log}) by its linearisation:
\begin{equation}
     \overline{\overline{U}}(\vec r) \approx  \frac{1}{2}\left( 
\frac{\overline{\overline{M}}(\vec r)}{M_0}-
     \overline{\overline{I}}
     \right).
     \label{def U}
\end{equation}

  \section{Results} \label{results}

\subsection{Transient regime}\label{transient}

\begin{figure}
\centering
\includegraphics[width=\linewidth]{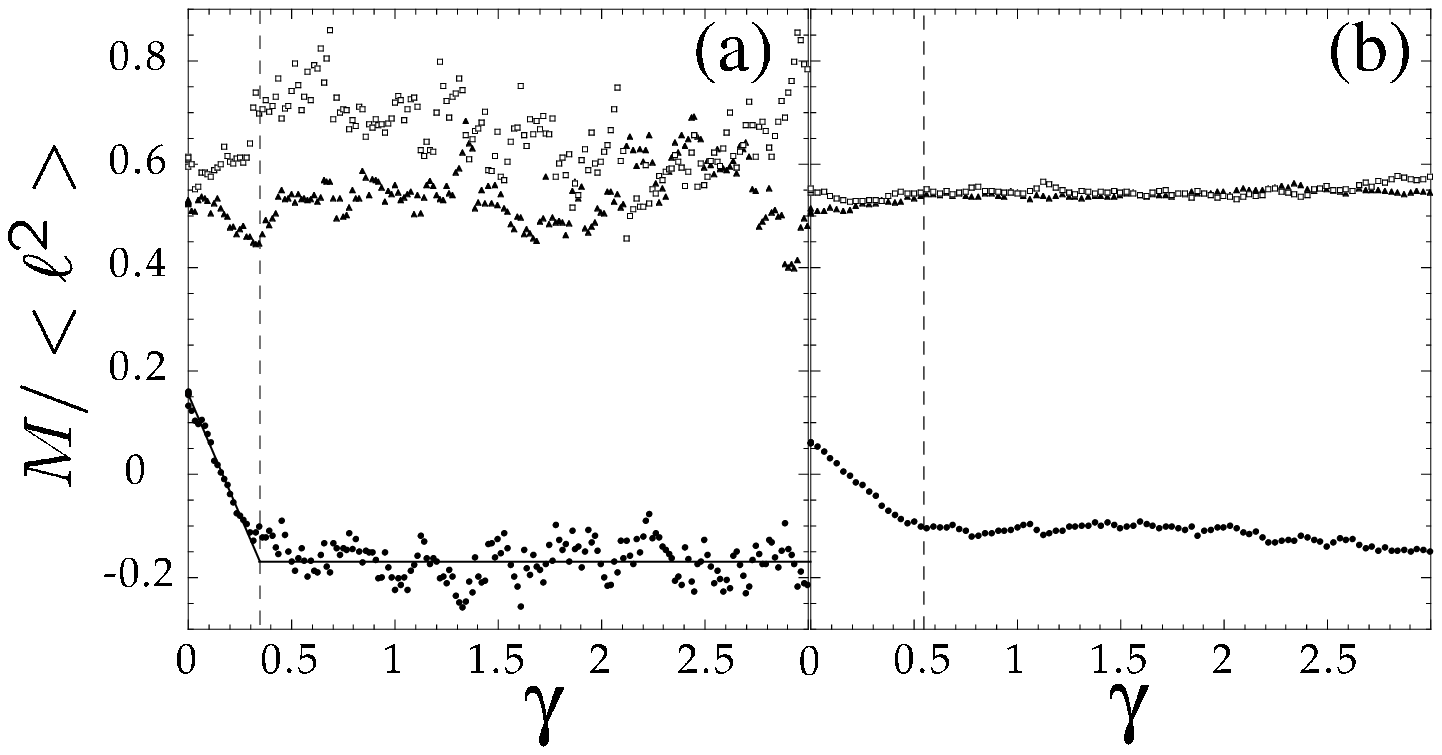}
\caption{Run 1 (transient regime):
texture tensor $\overline{\overline{M}} $
{\it versus} the applied shear strain $\gamma$.
The components
of   $\overline{\overline{M}} $
  defined in eq. (\protect\ref{defM}) and measured
from  the first 200 images
    ($\blacktriangle$: $M_{rr}$; $\square$: $M_{\theta\theta}$;
$\bullet$: $M_{r\theta}$)
are expressed in units of the average length of the bubble
    wall, $ \langle \ell^2 \rangle$.
(a) Innermost box (number 1):
each point represents one image;
(b) outermost box (number 8):
for clarity, we show only every second image.
The vertical dashed line marks the cross-over $\gamma_Y$ between the transient and
stationary regimes,  defined by the intersection between the 
straight line and the zero slope line fitting $M_{r\theta}$ in the transient 
and stationary regimes, respectively (solid lines in (a)).
}
\label{FigTransi}
\vspace{1cm}
\end{figure}

All data in this section are taken from Run 1 (Fig. \ref{images}),
which has two distinct regimes: a transient
and a stationary flow (Fig. \ref{FigTransi}).

\subsubsection{Texture tensor}

In the {\it transient regime}, the average wall length does not
change
much, but the walls tend to
align with the direction of the rotation (Fig. \ref{images}). Hence, 
it is mostly the cross-component $M_{r\theta}$ which correlates
to the shear  (Fig. \ref{FigTransi}). Since $\gamma=0$ marks the 
change from clockwise to
counter-clockwise shear,
$M_{r\theta}$ decreases linearly
and changes sign.
The correlation  with shear of both other components, if any, is of order of their fluctuations.

\begin{figure}
	\center
\includegraphics[width=9cm]{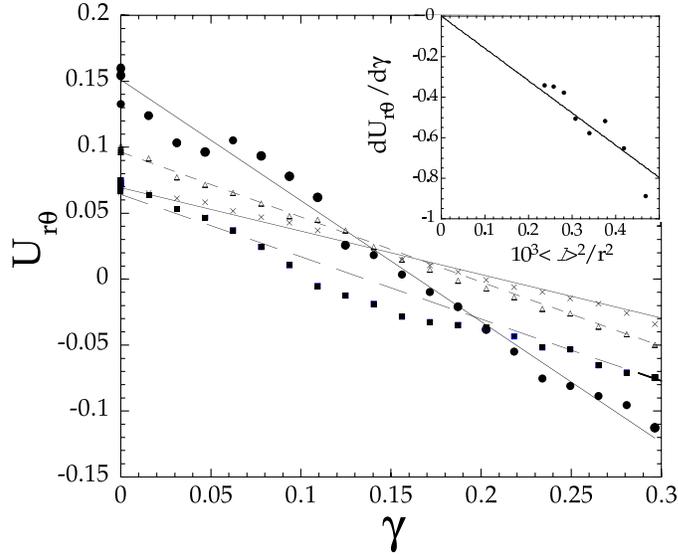}
\caption{Run 1 (transient regime):
cross component $U_{r \theta}$  {\it versus} the applied shear strain 
$\gamma$, for box 1: $\bullet$, box 3: $\blacksquare$, box 5: $\triangle$, and box 7: $\times$.   
Since $\gamma=0$ marks the reversal of the rotation
(and not an initially unstrained state), we expect $U_{r \theta}$ to be zero when $\gamma=0.15$,
within experimental error ($\pm$ 0.03). 
A linear fit of $U_{r \theta}$ {\it versus} $\gamma$ yields the derivative $d U_{r 
\theta}/d\gamma$, plotted in Insert, where 
the line is a linear  fit with zero intercept to  $d U_{r 
\theta}/d\gamma$
{\it versus} $1/r^2$.
	}
	\label{Urt lin}
\vspace{1cm}
\end{figure}

\subsubsection{Elastic strain tensor}

During the transient, no bubble rearrangement   (neighbour swapping, also called ``T1 
processes'') is visible for $\gamma < 0.3 $. This probably
corresponds to the elastic regime \cite{hohler}. 
 The cross-component  $U_{r \theta}$  varies linearly with $\gamma$, and
a linear fit yields its derivative  $d U_{r \theta}/d\gamma$
(Fig.  \ref{Urt lin}).
It  varies smoothly with $r$,  and a fit to  $r^{-2}$ 
yields the prefactor  $(- 1600 \pm 90)\: \left\langle\ell\right\rangle^2$.

It is interesting to compare  $d U_{r \theta}/d\gamma$
with the  expression which would be predicted by classical elasticity in a 
Couette geometry:
\begin{equation}
\frac{dU_{r\theta} }{d\gamma}  =  \frac{r_1^2 r_0}{r_1+r_0} \;  \frac{1}{r^2}.
\label{u couette}
\end{equation}
The prefactor of eq. (\ref{u couette}) is   
$ (r_1^2 r_0)/[(r_1+r_0)] = (- 1680 \pm 80)\: \left\langle\ell\right\rangle^2 $, 
where we have taken 1.2 mm (the depth of the teeth) as the uncertainty  of the wheel radius. 
This agrees with the experimental prefactor of $r^{-2}$
(Fig.  \ref{Urt lin}).

\subsubsection{Yield }\label{yield}

\begin{figure}
      \center
\includegraphics[width=8cm]{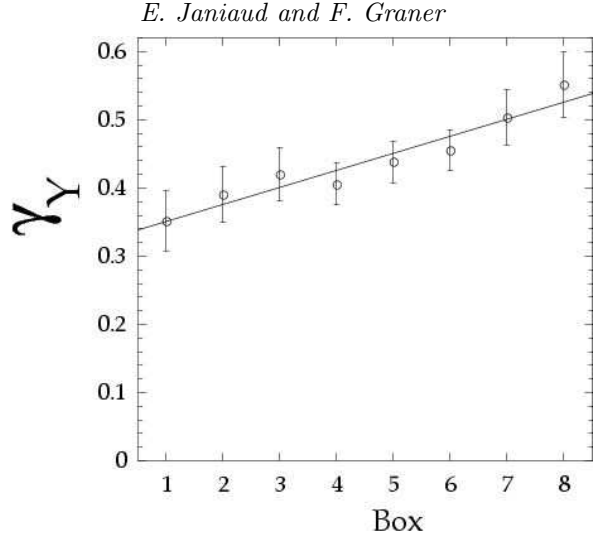}
  \caption{Run 1 (transient regime): variation of $\gamma_{Y}$
{\it versus} box number. Data are from   Fig. (\protect\ref{FigTransi}).
The line is a linear fit.
  }
  \label{Gamma s}
  \vspace{1cm}
\end{figure}

The flow becomes  {\it stationary } at a 
value  $\gamma_Y$ which we measure
by determining the intersection between  the transient build-up of  $M_{r 
\theta}$ fitted by a straight line and the 
stationary plateau  fitted by a horizontal line (Fig. \ref{FigTransi}).
It would be interesting to determine whether $\gamma_Y$ actually marks
the onset of   
irreversibility  in the stress-strain
relation at  macroscopic scale (plasticity).

We can at least  observe on the movie (data not shown) the 
  ``T1" 
bubble rearrangements, which  concern local topology.
In the inner boxes, 
$\gamma_Y$ 
correlates with the appearance of a large number of T1s.
The first isolated T1s
 appear  at values of $\gamma$ between 0.26 and 0.32, of the 
same order of magnitude as, but significantly smaller than $\gamma_Y$. 
The  outer boxes 
lack T1s, except once near $\gamma \sim 3$, well beyond $\gamma_Y$ and in the stationary regime.
Though 
$\gamma_Y$
is well defined, it marks a saturation in the  local shear strain applied to the bubbles,
which probably oscillates and remains below the actual yield strain.

The value of $\gamma_Y$
increases gradually
from
0.35 for the inner box, to 0.55 for the outer box
(Fig. \ref{FigTransi}). Within experimental error  ($\pm$ 0.03, corresponding to $\pm$ 2 images),
this variation is linear (Fig. \ref{Gamma s}).
   Explaining this variation of  $\gamma_Y$ with $r$  would certainly 
help  us to understand the yielding of foams (see section \ref{disc yield}).

\begin{figure}
\begin{center}
\includegraphics[width=9cm]{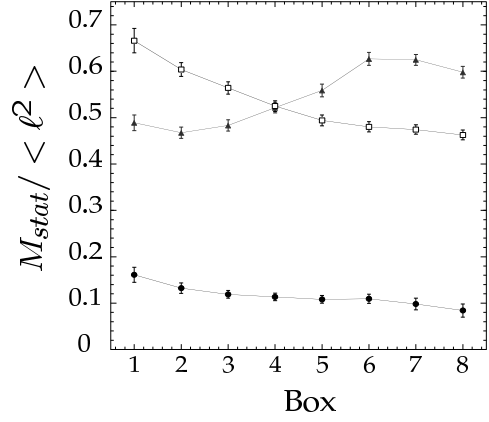}
\end{center}
\caption{
Run 2 (stationary flow): texture tensor 
$\overline{\overline{M}}_{stat} $ {\it versus} box number.
The components of $\overline{\overline{M}} $
($\blacktriangle$: $M_{rr}$; $\square$: $M_{\theta\theta}$;
$\bullet$: $M_{r\theta}$),
again expressed in units of $ \langle \ell^2 \rangle$, are
averaged over the four quarters of the whole foam,
and  also over time (2000 pictures).
}
\label{FigPerm}
\vspace{1cm}
\end{figure}

  \subsection{Stationary regime - average values}\label{permanent}

All data of this section are taken from  Run 2 (Fig. \ref{FigPhoto}),
which  allows us to improve our statistics, using temporal averages over
the stationary flow.

  \subsubsection{Texture tensor}

  Fig. (\ref{FigPerm}) plots  the average  $M_{\mathrm{stat}}$ of
   the instantaneous texture tensor $\overline{\overline{M}}$ in  the 
stationary regime.
We can now detect, not only the (small) anisotropy of the foam, 
$M_{r\theta}$, but  also 
its  small spatial   variations.

    $M_{r\theta}$ decreases across the gap: it is larger in the inner 
boxes than in the outer ones,
reflecting the preferred
orientation imposed by the shear.
    $M_{rr}$ decreases,
    $M_{\theta\theta}$ increases, and
   $M_{rr} - M_{\theta\theta}$ changes sign; this sign-change is more 
difficult to understand (section
\ref{nonzero}).

  \subsubsection{ Elastic strain}

\begin{figure}
	\center
	\includegraphics[width=9cm]{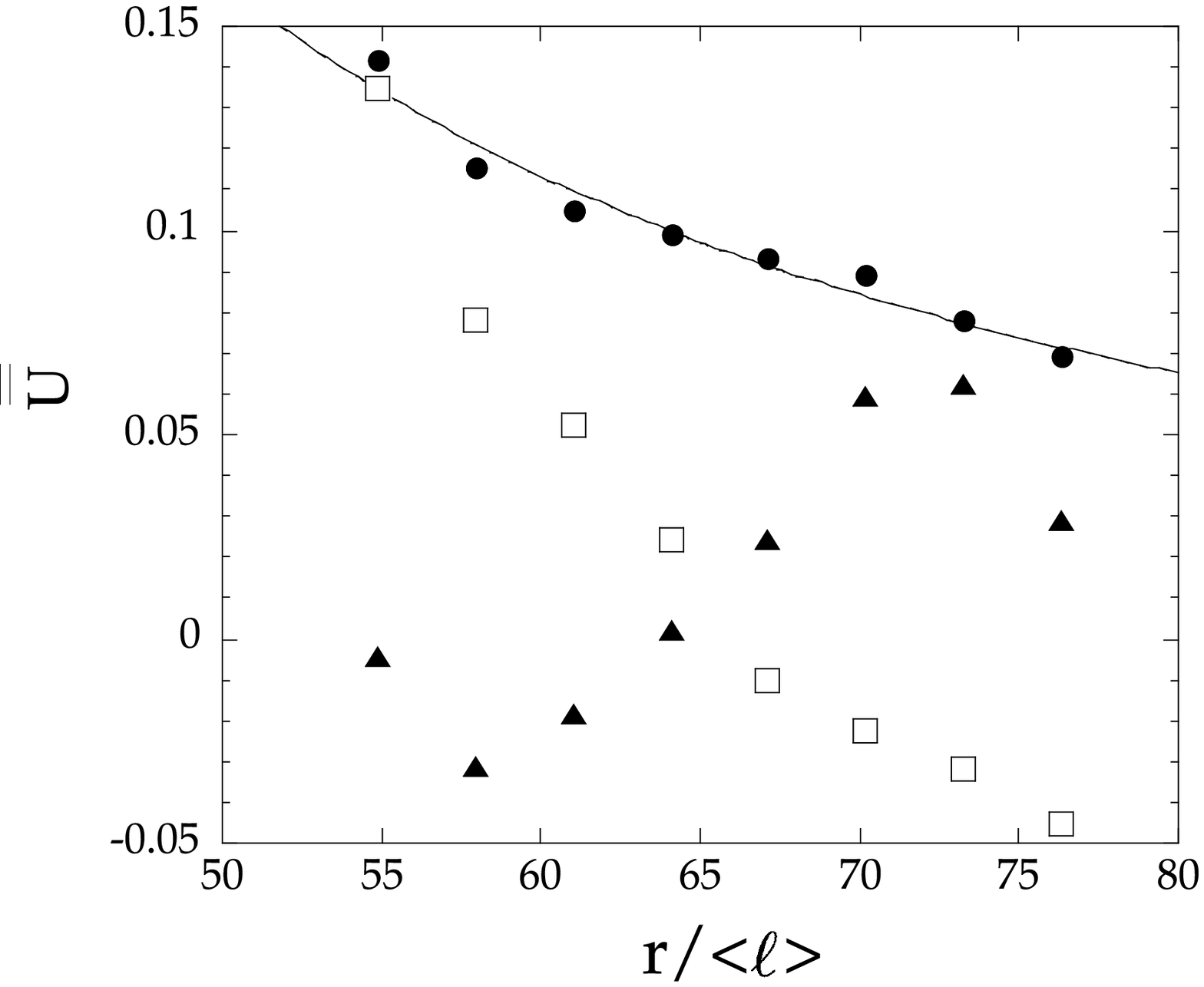}
\caption{
Run 2 (stationary flow): tensor $\overline{\overline{U}} $ {\it 
versus} box number.
The components of $\overline{\overline{U}} $
($\blacktriangle$: $U_{rr}$; $\square$: $U_{\theta\theta}$;
$\bullet$: $U_{r\theta}$)
are derived from the data in  Fig. (\protect\ref{FigPerm})
through eq.
(\protect\ref{def U}), using the value $M_0=1.0$ mm$^2$ (see section
\protect\ref{reference}).
The solid line is a best fit of   $U_{r\theta}$ to  a function
cst$/r^2$.
}
	\label{U(r)}
\vspace{1cm}
\end{figure}

  $U_{r\theta}$ does not localise
near the inner wheel: it varies smoothly  with the distance to the
inner wheel. It decreases as $\sim 1/r^{2}$ (Fig. \ref{U(r)}),
as expected if
the foam behaves like a continuous elastic medium, see eq. (\ref{u couette}).

To   compare  to  the transient regime in Run 1 
(which has slightly different bubble and wheel diameters),
we express the prefactor determined in Fig. (\ref{U(r)}) in the 
same units as in eq. (\ref{u couette}):
\begin{equation}
\langle U_{r\theta} \rangle= \gamma_{el} \frac{r_1^2 r_0}{r_1+r_0} 
\frac{1}{r^2}.
\label{gamma elast}
\end{equation}
The prefactor is $\gamma_{el} =  0.171 \pm 0.003 $,
see section (\ref{disc yield}).

\begin{figure}
	\center
	\includegraphics[width=9cm]{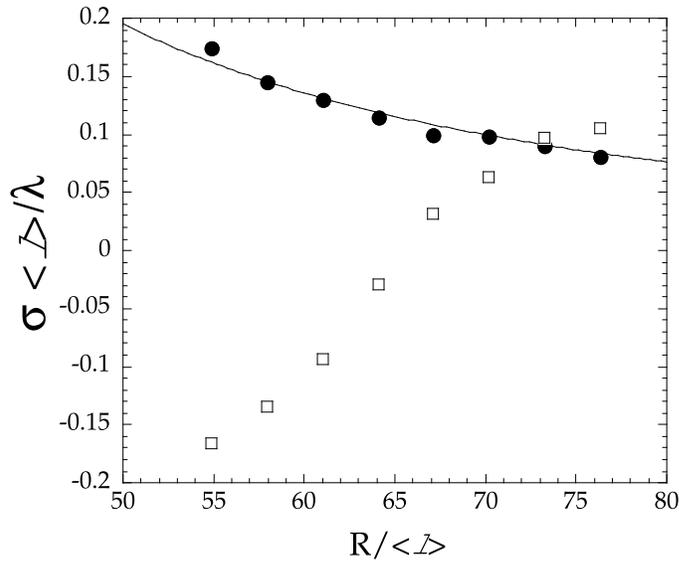}
	\caption{
Run 2 (stationary flow): elastic stress tensor 
$\overline{\overline{\sigma}}^{cap} $ {\it versus} box number.
The deviatoric components of $\overline{\overline{\sigma}}^{cap} $
($\bullet$: shear stress $\sigma^{cap}_{r\theta}$;
$\square$: normal stress difference 
$\sigma^{cap}_{rr}-\sigma^{cap}_{\theta\theta}$)
are derived from   eq.
(\protect\ref{sigma cap}) and expressed in units of  $\lambda/\langle 
\ell \rangle$.
The solid line is a best fit of $\sigma^{cap}_{r\theta}$ to  a function
cst$/r^2$.
	}
	\label{div stress}
\vspace{1cm}
\end{figure}

  \subsubsection{Deviatoric elastic stress} \label{deviatoric}

In the quasistatic regime, mechanical equilibrium implies that the 
elastic stress is divergence free.
Hence  we  expect that its off-diagonal component, which reduces to the capillary 
contribution, varies as for a continuous elastic medium in a Couette geometry 
\cite{dennin}:
$   \sigma^{cap}_{r\theta} \sim  {\mathrm 
cst}/r^2$ . Such $r^{-2}$ dependence occurs experimentally
(Fig. \ref{div stress}).

Again, the spatial variation of $\sigma^{cap}_{rr} - 
\sigma^{cap}_{\theta\theta}$, which changes sign,
is more difficult to understand (section
\ref{nonzero}).

  \subsubsection{Non-dimensional shear modulus} \label{hooke}

\begin{figure}
	\center
	\includegraphics[width=9cm]{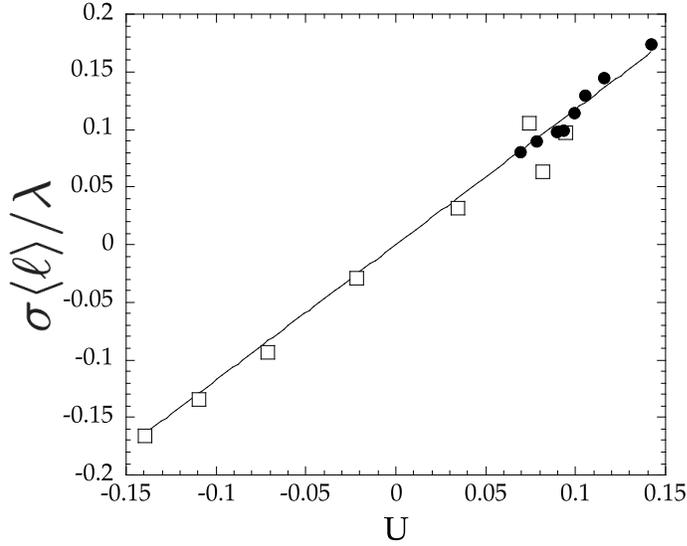}
	\caption{Run 2 (stationary flow):
Hooke relation.
The deviatoric part of the capillary stress 
$\overline{\overline{\sigma^{cap}}} $, expressed in units of 
$\lambda/\langle \ell \rangle$,
plotted {\it versus} the corresponding components of 
$\overline{\overline{U}} $
($\bullet$: $\sigma^{cap}_{r\theta}$ {\it versus}
$U_{r\theta}$; $\square$: 
$\sigma^{cap}_{rr}-\sigma^{cap}_{\theta\theta}$ {\it versus}
$U_{rr}-U_{\theta\theta}$).
Each data point comes  from   Figs.  
(\protect\ref{U(r)})    and  (\protect\ref{div stress}),
{\it i.e.}  it derives from averages at one position in the foam.
The solid line is a linear fit through all points, slope
	$2\mu \langle \ell \rangle/\lambda= 1.17 \pm 0.04$.
	}
	\label{Hook}
\vspace{1cm}
\end{figure}

In elasticity, the stress and the deformation are conceptually 
independent quantities (see section \ref{difference}). In  a 
continuous  medium, they 
correlate physically; their relation is the elastic part of the material's 
constitutive relation  \cite{landau}.
Let us focus on the components we measure here, namely the 
deviatoric terms $rr-\theta\theta$ and $r\theta$. If the material is 
linear, the components of strain and stress are proportional to each 
other, the slope being twice the shear modulus $\mu$. If the material 
is isotropic, $rr-\theta\theta$ and $r\theta$ play a similar  role. 
In polar coordinates, the Hooke equations becomes:

\begin{eqnarray}
\sigma^{cap}_{r \theta} &=& 2 \mu U_{r \theta} ,\\
     (\sigma^{cap}_{rr}-\sigma^{cap}_{\theta \theta}) &=& 2 \mu 
(U_{rr}- U_{\theta
     \theta}).
     \label{loi Hook}
\end{eqnarray}

Fig. (\ref{Hook})   indicates that, in the quasistatic regime, the foam 
indeed behaves like  a continuous medium:
the deviatoric part of the capillary stress 
$\overline{\overline{\sigma^{cap}}} $ and of $\overline{\overline{U}} 
$ correlate strongly, and details of the
microstructure appear only through mesoscopic averages.
Moreover, the foam is both  isotropic and linear, with a shear 
modulus $\mu= (0.59 \pm 0.02) \lambda / \langle \ell \rangle$. With a 
smaller or a larger box size, we obtain the same value for $\mu$ with a 
larger error.

  \subsection{Permanent regime - temporal fluctuations}\label{fluctuat}

\subsubsection{Time series of $\overline{\overline{M}}$ in Run 1}

The stationary flow results from a balance between the increase  due to 
the applied shear strain and relaxation due to bubble 
rearrangements \cite{langer}. The  texture tensor and the elastic strain  
fluctuate
around their mean value.

In Run 1, the short-term fluctuations are much larger near the inner 
wheel (Fig.
\ref{FigTransi}a),  where
shape deformations and bubble rearrangements  are both larger 
\cite{debregeas1},
than near the
outer wheel (Fig. \ref{FigTransi}b).
Fig.  (\ref{FigTransi}) does not show any significant physical 
feature of the fluctuation statistics, {\it e.g.}   abrupt, large-scale 
relaxations of deformations {\it via}
  correlated,  large stress drops.

\subsubsection{Histograms of increments in Run 2} \label{sec_histo}

\begin{figure}
	\center
	\includegraphics[width=0.9\linewidth]{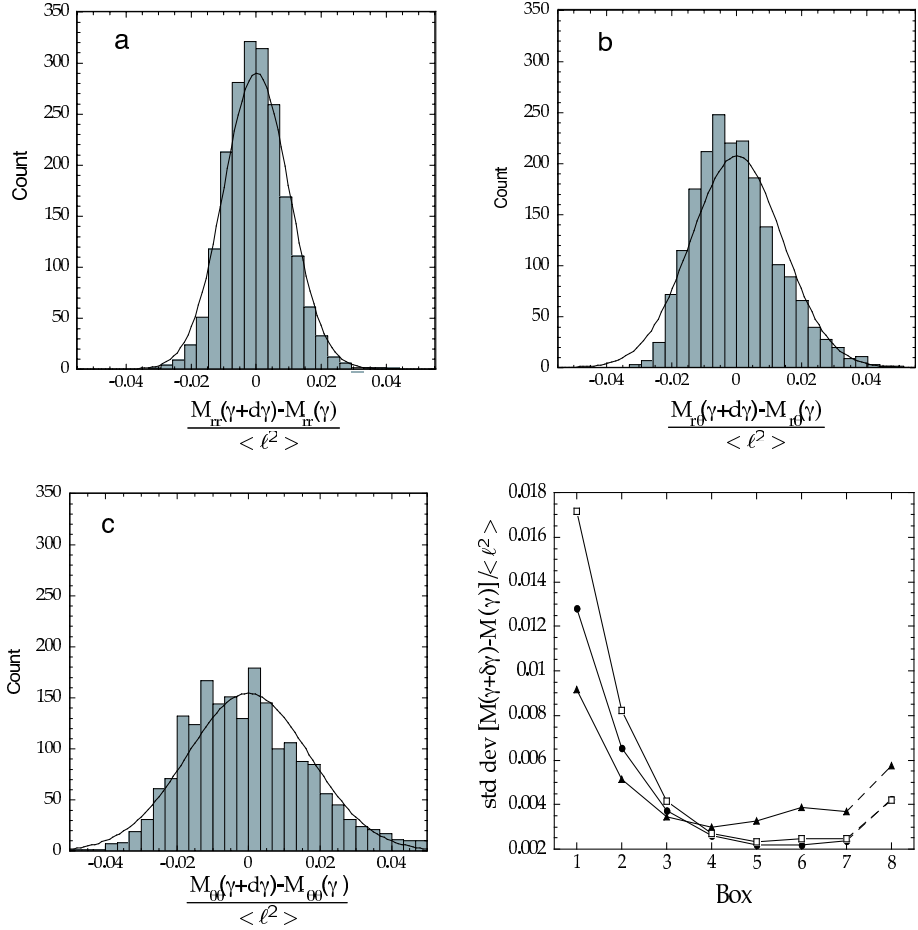}
	\caption{
Run 2 (stationary flow): statistics for the increments of the
texture tensor $\overline{\overline{M}} $.
  (a) Histogram of increments of $M_{rr}$
in box 1, again expressed in units of $ \langle \ell^2 \rangle$. Solid line
  is a Gaussian
with the same standard deviation and zero average.
(b, c) The same for
$M_{r \theta}$  and
	$M_{\theta \theta}$.
(d) Standard deviation of increments of $\overline{\overline{M}}$
{\it versus} box number ($\blacktriangle$: $M_{rr}$; $\square$: $M_{\theta\theta}$; $\bullet$: $M_{r\theta}$). 
Since the fluctuations in box 8 are extremely small (see
Fig. \protect\ref{FigTransi}), the data are noisy:  image filtering 
and thresholding would have been required to obtain a
statistically  significant 
value.
	}
	\label{fig_histo}
\vspace{1cm}
\end{figure}

We perform a  more quantitative analysis on Run 2, at a small scale, namely
between two successive images, corresponding to an increase of $d\gamma$ 
in the applied shear strain.
The increment of the texture tensor is defined as
$\overline{\overline{M}}(\gamma +d\gamma)-\overline{\overline{M}}(\gamma)$.
The signature of ``avalanche-like" events at such a small scale  would 
be extremely large fluctuations and/or  a large asymmetry between 
positive and negative increments.

We find no such signature.
The increments are almost Gaussian  (Fig. \ref{fig_histo}a-c).
The width of the histograms decreases by a factor of 10 from inner to 
outer boxes (Fig. \ref{fig_histo}d): as we mentioned, most large
fluctuations occur near the inner wheel.
Only  $M_{r\theta}$ displays a slight  asymmetry between positive and 
negative increments
  (Fig. \ref{fig_histo}b), probably because this cross-component is 
increased  by the applied shear strain and decreased by   the 
anisotropy relaxation due to T1 events.

  \subsubsection{Temporal autocorrelations in Run 2}

\begin{figure}
	\center
	\includegraphics[width=\linewidth]{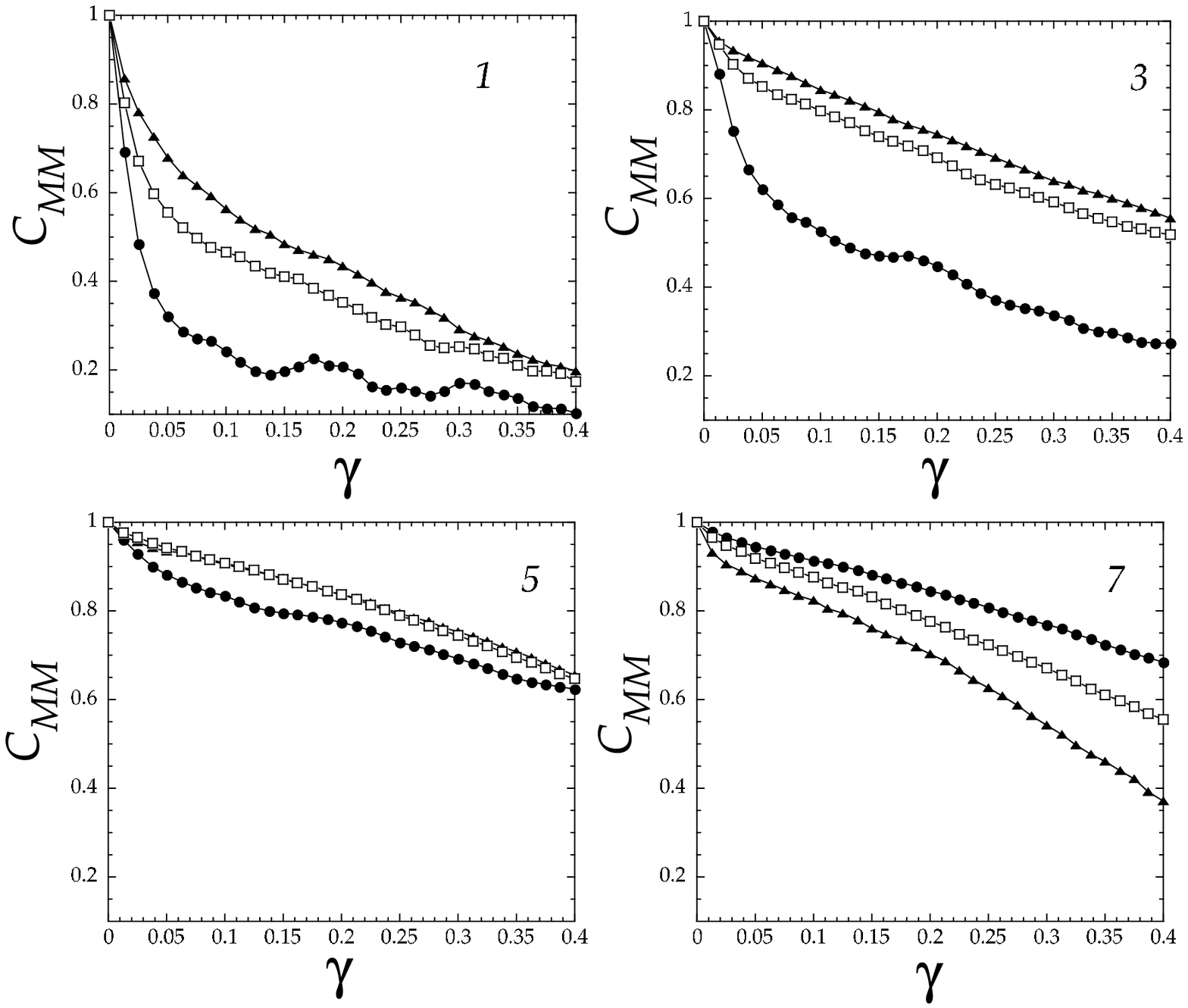}
	\caption{
Run 2 (stationary flow): temporal
autocorrelation $ C_{MM}$   of texture tensor
	$\overline{\overline{M}}$ (eq. \protect\ref{def auto}), {\it 
versus} $\delta{\gamma}$,  in boxes  1, 3, 5, 7
($\blacktriangle$: $M_{rr}$; $\square$: $M_{\theta\theta}$;
$\bullet$: $M_{r\theta}$).
We recall that the total applied shear $\gamma$ is proportional to time,
 not to the local shear in each box (eq. \ref{prop_time}).
	}
	\label{all autocor}
\vspace{1cm}
\end{figure}

The rotation of the inner wheel does not perturb the foam near the outer wheel.
In the outer boxes, the texture tensor thus decorrelates very slowly. Let us quantify this observation
using the data of Run 2.

To analyse the fluctuations at all scales of the shear strain $\delta 
\gamma$, we calculate the temporal autocorrelation  $ C_{MM}$ of the fluctuations:
\begin{equation}
  C_{MM}( \delta{\gamma})= \frac{
  \left\langle
  \left(M(\gamma)-M_{\mathrm{stat}}\right)
\left(M(\gamma+ \delta{\gamma})-M_{\mathrm{stat}}\right)
  \right\rangle}{
  \left\langle
  \left(M(\gamma)-M_{\mathrm{stat}}\right)^{2}
  \right\rangle},
  \label{def auto}
  \end{equation}
where 
 $M(\gamma)$  (resp: $M_{\mathrm{stat}}$)
is  the instantaneous value  (resp: the average over the stationary regime)
of the texture tensor $\overline{\overline{M}}$.

\begin{figure}
      \center
 \includegraphics[width=0.6\linewidth]{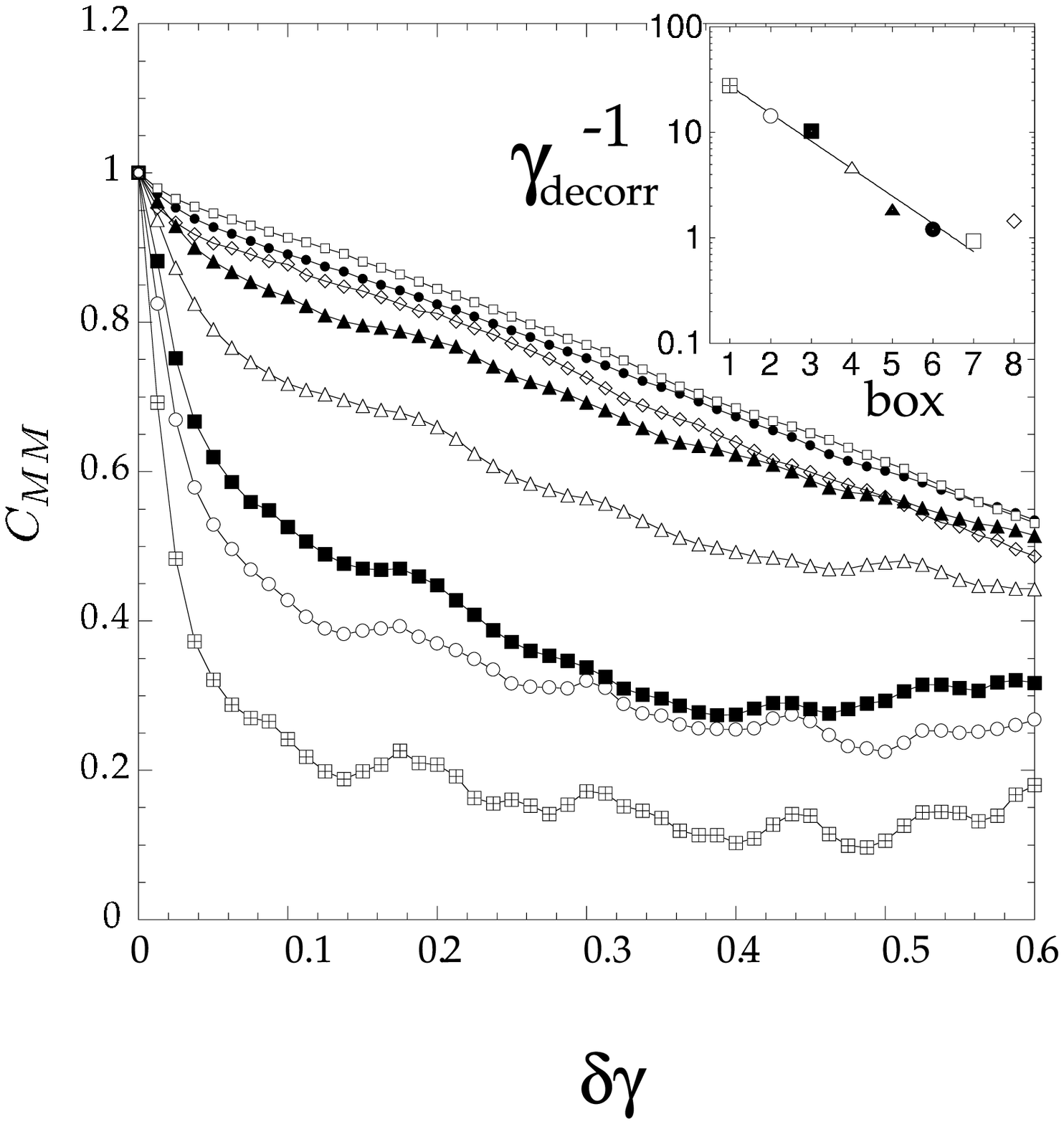}
  \caption{
Run 2 (stationary flow):
autocorrelation $C_{M_{r\theta}M_{r\theta}}$  of the cross-component
$M_{r\theta}$.
Box 1: $\boxplus$, box 2: $\bigcirc$, box 3: $\blacksquare$,
box 4: $ \triangle$, box 5: $\blacktriangle$, box 6: $\bullet$,
box 7: $\square$, box 8: $\Diamond$.
Insert:
initial decorrelation rate  $\gamma_{\mathrm{decorr}}^{-1}$
for  each box (eq. \protect\ref{def initial}).
}
  \label{autocor-M}
  \vspace{1cm}
\end{figure}

Fig. (\ref{all autocor}) displays  a rapid  initial decay in the correlation, 
followed by a slower decay,
reflecting the destruction  of the initial conditions \cite{durian,lauridsen,ybert,pratt}.
An exponential  fit at small  $\delta{\gamma}$ yields the
  shear strain $\gamma_{\mathrm{decorr}}$ characterising the 
initial 
  loss of correlation,  especially  quick  for  the cross 
component  $M_{r \theta}$ (Fig.
\ref{autocor-M}):
\begin{equation}
M_{r \theta} = \exp \left( -
\frac{\delta \gamma}{ 
\gamma_{\mathrm{decorr}}}\right).
\label{def initial}
\end{equation}
As we expect, the correlation decays much more quickly near the inner 
wheel, $\gamma_{\mathrm{decorr}}\approx 1/30 $, than near the outer 
one, $\gamma_{\mathrm{decorr}}\approx 1 $ (insert of Fig.
\ref{autocor-M}). 
The decorrelation rate $\gamma_{\mathrm{decorr}}^{-1}$ decreases exponentially
with the distance from  the inner wheel.

 The fluctuations of the 
texture tensor are thus  localised
near the inner wheel, over a distance comparable to 4  bubble diameters.
This  localisation resembles that of the 
velocity field over 1  bubble diameter
\cite{debregeas1}, and  is probably the signature of the 
effect of T1s on the deformation \cite{dennin}.

  \section{Discussion}\label{discu}

\subsection{Discussion of the strain and stress}\label{disc_stress_strain}


\subsubsection{Measurement method of the stress }

 Let us discuss 
the contributions to the stress we do not measure.

The dissipation is probably dominated by the 
friction on both glass plates, which we expect to be much larger than the
internal viscous stress \cite{cantat}.  
Whatever its physical origin, it
 should on average balance 
the input power. It is thus never negligible, especially during the 
relaxation after each T1, see section (\ref{qstat}).
However, the results we present here, which concern only the elastic 
terms, do not require us to measure the dissipation.

We measure the deviatoric elastic stress  using eq. (\ref{sigma 
cap}), which is formally equivalent to the stress in a network where sites 
interact through a two-body force. Foams do not include only
two-vertex interactions. Their stability is due to the 
pressure inside each bubble, which   relates to the bubble volume.
Hence pressure is a function of all wall curvatures and all the vertex 
positions. It is an $n$-body interaction, which we can decompose 
into two-point interactions in principle \cite{alexander}, but not 
really in practice.

It is not important here to measure the pressure inside bubbles, for the following reasons \cite{wuppertal}.
Since the pressure stress is 
isotropic, it does not contribute to
the deviatoric elastic stress. Hence the shear modulus is independent of the 
pressure field.
Finally,   the compression modulus of a foam is typically that of
an ideal gas, which scales like the pressure inside bubbles $\sim 10^5$ Pa
(multiplied by $h$, when expressed as a 2D compression modulus).
This modulus is not very interesting  and  is much larger 
than the shear modulus  of  the foam, which is thus effectively  incompressible 
\cite{hutzler}.


The line tension is not known, but unimportant for our present  
purposes. The stress, and hence the shear 
modulus, are expressed in non-dimensional  form, as
required for comparison to other foams. A direct measurement of $\lambda$ 
  can complement
 the local image 
measurement   
when necessary,  for instance to
predict the actual
  force exerted by
the flowing foam on an obstacle, or to validate
eq. (\ref{sigma cap}) by checking 
its agreement  with  an independent,
macroscopic measurement of the force  \cite{langmuir_foam}.

\subsubsection{Difference between 
$\overline{\overline{\sigma}}^{cap}$  and  $\overline{\overline{M}} 
$} \label{difference}

Both $\overline{\overline{\sigma}}^{cap}$  and  $\overline{\overline{M}} $
are measured from  the same image, and their definitions  
(eqs. \ref{sigma cap}, \ref{defM})
seem similar. However, they are really independent 
both physically and mathematically.

Physically, their significance and status are different.
The value of  $\overline{\overline{\sigma}}^{cap}$    can be 
determined from image analysis only if the sites interact through a 
two-body force $\vec{\tau}$, and if we explicitly know the 
form of this force $ \vec{\tau}(\vec{\ell})$    with respect to 
the distance between sites.
The value of $\overline{\overline{\sigma}}^{cap}$ is sensitive  to physical forces:
first, it doubles if the 
density of links   doubles; second, it  does not  change  if we cut  a link in two by artificially 
introducing a new site at its middle; third, in  foams,
it formally depends on wall curvatures.

On the other
hand,
$\overline{\overline{M}} $  is purely geometric.
  It is a characteristic of the pattern, independent of   physical 
interactions; it can be defined and measured for  various kinds of 
cellular patterns for which  no $\vec{\tau}$ can be defined, including 
solid foams, and even  grain boundaries in crystals 
\cite{durand_weiss}. 
The  value of  $\overline{\overline{M}} $   is sensitive   to the  detailed network 
topology:
first, it does not  change if the 
density of walls   doubles; second, it decreases if we cut  a wall in  two by artificially 
introducing a new vertex at its middle; third, in foams, it  is insensitive to the (in-plane or 
out-of-plane) curvature of walls. 

Mathematically, $\overline{\overline{\sigma}}^{cap}$  and 
$\overline{\overline{M}} $
 are independent in all  cases, except when  $\vec{\tau} \propto
\vec{\ell}$, {\it i.e.}  a harmonic spring, with a length 
$\ell$ much greater than its length at equilibrium. For instance, writing
  the stress 
 of  an 
isolated polymer molecule in a shear solvant  as $\vec{\ell} \otimes \vec{\ell}$ 
 is legitimate  \cite{bird}. 
However,  foams, for which $ 
\tau = $ const, fall in the general case where
  $\overline{\overline{\sigma}}^{cap}$  and  $\overline{\overline{M}} 
$  are mathematically different averages.
The difference is even more visible in a 3D foam, where 
$\overline{\overline{\sigma}}^{cap}$
is an integral over a surface \cite{batchelor}, while 
$\overline{\overline{M}} $  remains based on vectors
\cite{aubouy}.  

Moreover, $\overline{\overline{M}} $ does not even need be defined on the vectors $\vec{\ell}$ linking neighbour vertices. Definitions other than eq. (\ref{defM}) could be acceptable in principle \cite{dollet}. For instance, we could use the vector $\vec{g}$ which links the centers of mass of neighbouring bubbles, and construct $\overline{\overline{M}}$ as $
\left\langle f\; \vec{g}\otimes \vec{g} \right\rangle$. This would apply to bubble rafts \cite{dennin}, granular materials \cite{kruyt}, or 3D foams \cite{kraynik2003}.
This would  
change both $\overline{\overline{M}} $
and $\overline{\overline{M}}_0 $ by a multiplicative constant:
the elastic strain $\overline{\overline{U}} $ 
(eqs. \ref{U log}, \ref{def U}), 
and hence the shear modulus  $\mu$ (eq.     \ref{loi Hook})
would almost not change. 

We emphasise that $\overline{\overline{\sigma}}^{cap}$  and 
$\overline{\overline{U}} $  are  also  independent.
They represent different averages of 
microscopic details,  so we cannot  express one as a 
function of the other \cite{kruyt,gold3}, except empirically, 
 through numerical simulations
or physical approximations
approximations 
\cite{kruyt}. 
The difference is even more visible 
at large deformations: 
the  logarithmic term of eq. (\ref{U log}) 
deviates from the linear approximation when the maximum value of the elastic shear 
strain (0.15 in Fig. \ref{U(r)}) reaches large  values, {\it e.g.} 
0.4 \cite{langmuir_foam} or 
0.6  \cite{asipauskas}.
There is no trivial relation between them, except in a regular pattern, where averages are replaced by exact identities; for instance, in the case of a honeycomb, their relation is exactly as expected \cite{dollet}.

\subsubsection{Sign of normal differences}\label{nonzero}

The normal differences  $M_{rr}-M_{\theta\theta}$
and $\sigma^{cap}_{rr}-\sigma^{cap}_{\theta\theta}$ 
change sign in the middle of the gap,
between boxes 4 and 5 (Figs.  \ref{FigPerm}, \ref {U(r)},  \ref{div stress}).

Normal stresses appear in 
elastic solid \cite{poynting} or granular \cite{bagnold} materials sheared at constant volume
(this is equivalent to ``dilatancy": volume increase under shear at constant pressure  \cite{dilatancy}).
In complex fluids such as foams, these normal stresses exist but are difficult to measure \cite{khan1988} and to predict  \cite{dilatancy}.
Predictions and 2D simulations of their average are accurate: under steady shear, the pressure  of a foam with  fluid fraction $\Phi= 5\%$ increases by 15\%  \cite{dilatancy}.  This is the order of magnitude of the spatial variations  
 of 
 $M_{rr}$ and 
$M_{\theta\theta}$ (Fig.  \ref{FigPerm}).
We can not predict
their anisotropy (hence the normal stress differences), but normal stresses due to dilatancy are a second-order effect in the strain \cite{larson}: they are likely smaller that the large value of  $M_{rr}-M_{\theta\theta}$ we observe, and 
might be masked by dynamical effects due to bubble rearrangements.
 But we  can imagine other explanations.

First, this change could reflect an initial anisotropy resulting from  the 
procedure used to fill the cell
with foam \cite{debregeas1},
which the preparatory rotation does not suffice to relax  in the outer boxes
(Debr\'egeas, personal communication). The initial state is not necessarily 
a zero-stress state.
More generally, trapped stresses make notoriously difficult to measure
normal stress differences
\cite{khan1988}, and carefully relaxed simulations are more reliable  \cite{kraynik2003}. The present measurement is probably one of the first direct experimental  ones,
along with that of Ref. \cite{labiausse}.

Second, it could be an artefact due to our measurement method, which 
does not accounts for the out-of-plane curvature of bubble walls 
(in the third dimension). Since the foam is not perfectly dry, the sides 
of a wall can have slightly different curvatures, so that their contributions would 
not exactly cancel out.

Third, this change could arise from a deviation from axisymmetric Couette flow,
due to a secondary recirculation
with radial velocity components.
A signature of the discrete and disordered nature of the material is the 
appearance of  solid body rotation
of bubble clusters. It results  in vortices,
generic for amorphous elastic materials,
generated by the ``backflow" of the non-affine displacements
 \cite{tanguy}.
Their  lifetime is as 
short as the time the internal wheel takes to move one bubble diameter 
\cite{debregeas1}.

\subsection{Discussion of the stationary flow}

The present measurements clarify a few open debates.

\subsubsection{To localise or not to localise?}

The experimental set-up, the fluid fraction,  the friction on the plates, and the  boundary conditions at 
the wheel, are those of Ref. \cite{debregeas1}.
How can the same data alternatively display and not display  localisation?

There is no fundamental incompatibility:  different fields
have different spatial variations.
 Ref.  \cite{debregeas1} 
measures the velocity field, and the T1 rate; while we measure the  elastic 
stress and strain fields.
 In   a quasistatic regime, where the 
foam is almost always close to mechanical equilibrium, we expect 
the  stress  to vary smoothly across the gap, as observed here and in 
simulations \cite{debregeas2}.
The elastic strain, which correlates to the 
elastic stress,  does not localise, although the plastic strain,
which correlates to the velocity gradient,  does.

\subsubsection{Yield}\label{disc yield}

   The value of the yield strain and stress, characterised {\it e.g.} by the 
equality of the real and imaginary parts of the complex shear modulus
\cite{asnacios},
 is not completely understood yet. 
It generally  depends on the shear rate   
\cite{coussot,rouyer}, but not in the present quasistatic regime.

We expect that, to store a deformation $ 
\gamma_{el}$, requires us to apply an equivalent shear strain 
$\gamma_Y $.
Since Run 1 corresponds to an inversion from counter-clockwise to 
clockwise rotation, we expect that  $\gamma_Y = 2 \gamma_{el}$. 
We observe an excellent agreement
 in box 1, in which $\gamma_Y=0.35  \pm 0.02$
(Fig. \ref{FigTransi})
and $\gamma_{el}= 0.171 \pm 0.003$ (Fig. (\ref{U(r)}) and eq.
(\ref{gamma elast})).
But  $\gamma_Y \neq 2 \gamma_{el}$ in other boxes, for larger  $\gamma_Y$ 
(Fig. \ref{Gamma s}). 
Thus the deformation stored in the foam changes with $\gamma$ even
above 
 $\gamma_Y$. Kabla and Debr\'egeas
reach the same conclusion: their simulations indicate that, after the 
foam yields,
T1s localise in the shear band only after a certain delay time 
\cite{debregeas2}.

\subsubsection{Averages and fluctuations}

Our data show that stress and strain have a   well defined, physically  meaningful average
 at the scale of a representative volume element,
while their  fluctuations remain small.
A coarse-grained    shear modulus can be defined and measured,
unlike {\it e.g.}  in granular 
media, where the importance of fluctuations
at large scales
is still an open debate \cite{gold4}. 

More precisely, the statistics of the texture tensor are nearly  Gaussian. 
Visual inspection of the images confirms that T1s remain isolated. They
correlate \cite{debregeas2}, but over a very 
small space and time range, probably due to the experimental conditions: disordered foam, 
quasistatic regime, large dissipation due to top and bottom plates.
We do not observe
extremely large fluctuations, nor sudden relaxations of deformations and stress. 

Averages  seem to 
dominate the physics
Fig.  (\ref{FigTransi}), but fluctuations too are important. 
For instance, the width of the histogram of the velocity  or stress field fluctuations 
might  constrain the velocity field, and explain why 
shear-banding occurs preferentially near the inner wheel, due to a 
feedback loop on which  fluctuations grow  \cite{debregeas2}.

Both the 
average 
and fluctuations of T1 rate manifest themselves in the texture tensor.
The signature of the
average T1
rate  is the existence of a stationary regime, where the decrease in deformation due to T1s balances
the increase due to applied shear.
The  signature of T1 rate fluctuations is the 
exponential localisation of the texture tensor fluctuations.

\subsubsection{ Measurement of the shear modulus} \label{comments modulus}

Fig. (\ref{Hook})  suggests a few remarks.
First, the classical method to measure a shear modulus consists in increasingly 
deforming a sample, and measure the stress and strain in successive states.
We use here another  method  \cite{asipauskas}.
Each point in Fig. (\ref{Hook})  comes from a different
region of the same foam.  We thus observe different deformation states 
 simultaneously from  a single 
heterogeneous  image. 
Averaging over successive images reduces noise \cite{asipauskas}, but is not 
necessary  \cite{langmuir_foam,durand_weiss}. 

Second,  when determining the constitutive relation for plastic materials,  the  elastic stress 
is classically plotted {\it versus } the
total applied shear strain: see Refs. \cite{jiangpre,kraynik2000,dilatancy} to quote a few.
Different applied strains correspond to the same stress.
Here we  observe a one to one correspondance
between elastic stress and elastic strain (Fig.  \ref{Hook}).
The 
foam is elastic,  although the plastic strain increases steadily, and the foam 
is well beyond the yield strain.  

Third, although we do not know the value of $\lambda$, we can compare 
our non-dimensional measurement of shear modulus $\mu$  with  published  values.
An ideal two-dimensional dry monodisperse foam, where  bubbles form a 
regular  honeycomb lattice
\cite{princen,khan1986,hutzler}, has a shear modulus
  $\mu_{hc}=\frac{\lambda}{\sqrt{3} \langle \ell \rangle}
=0.577\frac{\lambda}{\langle \ell \rangle} $.
Our  $\mu$ is slightly larger, probably due to disorder; here  $\frac{\langle \ell^2 
\rangle}{\langle \ell \rangle^2}=1.08$. The non-zero fluid 
fraction  tends to decrease $\mu$ \cite{princen,khan1986,mason,dilatancy}.
In a non-quasistatic, and more heterogeneous, stationary flow,  either through a 
constriction or around an obstacle, a dry foam (fluid fraction much 
lower than $5\%$), even with a small bubble size dispersity  $<5\%$,
displays a shear modulus 20\% higher than 
$\mu_{hc}$, due  to
the stretched bubble shape ($\frac{\langle \ell^2 \rangle}{\langle \ell
\rangle^2}=1.22$
\cite{asipauskas}.

  \section{Conclusion}

To summarise, we re-analyse the data generously provided by Georges 
Debr\'egeas on a foam sheared in a two-dimensional Couette geometry,  with 
  the velocity field localised near the inner wheel 
\cite{debregeas1}. Since the flow is  quasistatic, the 
deformation of the bubbles is very small. We take advantage of the foam's
invariance under rotation
(and, in the stationary regime, of its time invariance) to 
average over many bubbles. We then measure the texture tensor of the foam, a general tool to 
quantify the deformation of 2D or 3D networks and cellular patterns 
\cite{aubouy}.

For small applied shear strain,   a transient regime has the 
cross-component  of the texture tensor increase linearly with shear,
with no  bubble rearrangements (``T1"), and the foam seems to 
behave like a classical elastic, continuous medium. 
The 
only large scale manifestation of the discrete, cellular nature of 
the foam is that, in 
 a mixture of 
two fluids (soap solution and air),
the surface tension of bubble walls produces
 elastic behaviour and a shear modulus.

The deformation saturates
first near the inner wheel, then gradually in more distant boxes,
at an applied shear strain $\gamma_{Y}$.
This onset of plastic behaviour correlates, but does not strictly coincide
with the appearance of T1s.

The foam then enters a stationary regime, where the deformation increase
 due to 
shear and the relaxation due to T1s balance each other. The 
fluctuations of the texture tensor around its mean  value 
remain small.
Despite the T1s, the elastic behaviour of the foam again resembles 
that of a classical elastic, continuous medium.  The 
maximum shear strain that the foam can sustain in the quasistatic regime 
is $\gamma_{el}$, compatible with the value of $\gamma_{Y}$. 
The 
strain and stress are highly correlated, even proportional 
to each other; the shear modulus is slightly higher than for  a 
honeycomb with similar average bubble wall length.
The deviatoric components of the elastic strain and stress tensors, 
as well as the texture tensor, do not localise.
The only trace of localisation is the  much quicker loss of correlation of the 
texture tensor fluctuations  near the inner wheel.

Future studies  aim at  understanding the origin of the value of the yield strain, 
the change of sign in the normal stress and strain differences, the 
precise role of boundary conditions, and the relation  to granular 
materials.
They  include the measures of pressure and dissipation, and the 
analytical calculation of the shear modulus.

 {\bf Acknowledgments:} 
We are extremely grateful to G. Debr\'egeas for providing us with
many recordings of his published and unpublished
experiments, and for detailed discussions. We would also like to thank
M. Dennin, R. H\"ohler, A. Kabla, A. Kraynik, N. Kruyt, J. Scheibert for
discussions, 
M. Aubouy,  B. Dollet, F. Elias and J.A. Glazier for critical reading of the manuscript,
O. Cardoso for help with image analysis. This
work was partially
supported by CNRS ATIP 0693.



\begin{thebibliography}{0}



\bibitem[(Alexander  1998)]{alexander}
{\sc Alexander S.}
{1998},
{\it Phys. Reports}, {\bf 296}, {65}.

\bibitem[(Asipauskas  {\it et al.}  2003)]{asipauskas}
{\sc Asipauskas M., Aubouy M., Jiang Y.,  Graner F.
\& Glazier J.A.}
{2003},
    {\it Granular Matt.}, {\bf 5}, {71}.

\bibitem[(Aubouy  {\it et al.}  2003)]{aubouy}
{\sc Aubouy M., Jiang Y., Glazier J. A. \& Graner F.}
{2003},
    {\it Granular Matt.}, {\bf  5}, {67}.

\bibitem[(Bagnold 1941)]{bagnold}
{\sc Bagnold R.A.} {1941},
{\it The Physics of Blown Sand and Desert Dunes}, Methuen, London.

\bibitem[(Ball   \& Blumenfeld   2002)]{ball}
{\sc Ball R. \& Blumenfeld R.}
{2002},
{\it Phys. Rev. Lett.}, {\bf 88}, { 115505}.

\bibitem[(Batchelor 1970)]{batchelor}
{\sc Batchelor G.K.} {1970}, {\it J. Fluid Mech.}, {\bf 41}, {545}.

\bibitem[(Bird, Armstrong \& Hassager 1987)]{bird}
{\sc Bird R.B., Armstrong R.C. \& Hassager O.} {1987},
{\it Dynamics of polymeric liquids},
John Wiley \& Sons, NY (USA).

\bibitem[(Blumenfeld   2004)]{blumenfeld}
{\sc Blumenfeld R.} 
{2004},
{\it Physica A}, {\bf 336}, { 361}.

\bibitem[(Cantat, Kern  \& Delannay 2004)]{cantat}
{\sc Cantat I., Kern N. \& Delannay R.} {2004},
{\it Europhys. Lett.} {\bf 65}, {726}.

\bibitem[(Courty {\it et al.}  2003)]{langmuir_foam}
{\sc  Courty S., Dollet B., Elias F., Heinig P. \& Graner F.}
{2003},
    {\it  Europhys. Lett.} {\bf 64}, {709}.

\bibitem[(Coussot {\it et al.} 2002)]{coussot}
{\sc
Coussot P.,  Raynaud J.-S., Bertrand F.,  Moucheront P.,  Guilbaud 
J.-P.,  Huynh H.T.,  Jarny S.  \&  Lesueur D. } {2002},
{\it  Phys. Rev. Lett. } {\bf 88}, { 218301}.

\bibitem[(Cox, Weaire \& Vaz  2002)]{Cox}
{\sc  Cox S.J., Weaire D., \& Vaz M.F.}
{2002},
{\it Eur. Phys. J. E}, {\bf 7}, {311}.

\bibitem[(Debr\'egeas, Tabuteau  \& Di Meglio  2001)]{debregeas1}
{\sc Debr\'egeas G., Tabuteau H. \& Di Meglio J.-M.}
   {2001},
  {\it Phys. Rev. Lett.}, {\bf 87}, {178305}.

\bibitem[(Dennin 2004)]{dennin}
{\sc Dennin M.} {2004},

\bibitem[(Dollet {\it et al.} 2004)]{dollet}
{\sc Dollet B., Mecke K., Aubouy A. and Graner F.} {2004},
in preparation.

\bibitem[(Durand, Weiss  \& Graner   2004)]{durand_weiss}
{\sc   Durand G., Weiss  J. \& Graner  F.}
{2004},
{\it Europhys. Lett.}, {\bf 67}, {1038}.

\bibitem[(Durian  1997)]{durian}
{\sc Durian D. J.}
   {1997},
  {\it Phys. Rev. E}, {\bf 55}, {1739}.

\bibitem[(Elias  {\it et al.}  1999)]{elias}
{\sc Elias F., Flament C., Glazier J. A., Graner F. \&
Jiang Y. }
{1999}, {\it Philos. Mag. B}, {\bf 79}, {729}.

\bibitem[(Farahani \& Naghdabadi 2000)]{farahani}
{\sc Farahani K. \& Naghdabadi R.} {2000},
{\it Int. J. Solids Struct.}, {\bf 37}, {5247}.

\bibitem[(G\'eminard {\it et al.}  2004)]{geminard}
{\sc G\'eminard J.-C.,  {\.{Z}}ywoci\'{n}ski A., Caillier F. \& Oswald P.} {2004},
{\it Phil. Mag. Lett.} {\bf 84}, {199}.

\bibitem[(Goldhirsch \& Goldenberg 2002)]{gold3}
{\sc Goldhirsch I. \& Goldenberg C.} {2002},
{\it Eur. Phys. J. E}, {\bf 9}, {245}.

\bibitem[(Goldhirsch \& Goldenberg 2004)]{gold4}
{\sc Goldhirsch I. \& Goldenberg C.} {2004},
{\it Granular Matter}, {\bf 6}, {87}.

\bibitem[(Graner {\it et~al.} 2001)]{graner2001}
{\sc Graner F., Jiang Y., Janiaud E. \&  Flament C.} 
{2001},
  {\it {Phys. Rev. E}},  {\bf{63}},  {011402}.

\bibitem[(Graner 2002)]{wuppertal}
{\sc Graner F.} {2002},
{\it Two-dimensional fluid foams at equilibrium}, in Ref. 
\protect\cite{mecke}, pp. 187-214.

\bibitem[(Hoger 1987)]{hoger}
{\sc Hoger A.} {1987},
{\it Int. J. Solids Struct.}, {\bf 23}, {1645}.

\bibitem[(H\"{o}hler, Cohen-Addad \& Asnacios  1999)]{asnacios}
{\sc H\"{o}hler R., Cohen-Addad S. \& Asnacios  A.}
    {1999},
{\it Europhys. Lett.}, {\bf 48 }, {93}.

\bibitem[(H\"{o}hler, Cohen-Addad \& Hoballah  1997)]{hohler}
{\sc H\"{o}hler R., Cohen-Addad S. \& Hoballah H.}
    {1997},
{\it Phys. Rev. Lett.}, {\bf 79}, {1154}.

\bibitem[(Jiang  {\it et al.}  2000)]{delft}
{\sc Jiang Y., Asipauskas M., Glazier J. A., Aubouy M. \& Graner F.}
{2000}.
In {\it Eurofoam 2000}  ed. Zitha P.,  Banhart J. \& Verbist  G., p. 
297, MIT Verlag,
Bremen.

\bibitem[(Jiang {\it et al.}  1999)]{jiangpre}
{\sc  Jiang Y., Swart P.J., Saxena A., Asipauskas M.
     \& Glazier J.A.}
{1999},
{\it Phys. Rev. E}, {\bf 59}, {5819}.

\bibitem[(Kabla   \& Debr\'egeas  2003)]{debregeas2}
{\sc  Kabla A. \& Debr\'egeas G.}
   {2003},
  {\it Phys. Rev. Lett.}, {\bf 90}, {258303}.

\bibitem[(Kern \& Weaire 2003)]{kern}
{\sc Kern N. \& Weaire D.} {2003},
{\it Phil. Mag. } {\bf 83},{ 2973}.

\bibitem[(Khan \& Armstrong 1986)]{khan1986}
{\sc  Khan S.A. \& Armstrong R.C.} {1986}, 
{\it J. Non-Newt. Fluid 
Mech.}, {\bf 22}, {1}.

\bibitem[(Khan \& Armstrong 1987)]{khan1987}
{\sc  Khan S.A. \& Armstrong R.C.} {1987}, {\it J. Non-Newt. Fluid 
Mech.}, {\bf 25}, {61}.

\bibitem[(Khan, Schnepper  \& Armstrong 1988)]{khan1988}
{\sc  Khan S.A., Schnepper  C.A. \& Armstrong R.C.} {1988}, 
{\it J. Rheol.}, {\bf 32}, {69}.

\bibitem[(Kraynik, Reinelt  \& van Swol  2003)]{kraynik2003}
{\sc  Kraynik A., Reinelt D., \& van Swol F.}
{2003},
{\it Phys. Rev. E}, {\bf 67}, { 031403}.

\bibitem[(Kruyt 2003)]{kruyt}
{\sc Kruyt N.P.} {2003},
{\it Int. J. Solids Struct.}, {\bf 40}, {511}.

\bibitem[(Kruyt \& Rothenburg 1996)]{kr96}
{\sc Kruyt N.P. \& Rothenburg L.} {1996},
{\it J. Appl. Mech.}, {\bf 118}, {706}.

\bibitem[(Kruyt \& Rothenburg 2002)]{kr02}
{\sc Kruyt N.P. \& Rothenburg L.} {2002},
{\it Int. J. Solids Struct.}, {\bf 39}, {311}.

\bibitem[(Labiausse, H\"ohler \& Cohen-Addad 2004)]{labiausse}
{\sc Labiausse V., H\"ohler  R. \& Cohen-Addad S.} {2004}
in {\it Proceedings of Eufoam 2004}, {\sc Adler M.} ed., Elsevier, to appear.

\bibitem[(Landau \&  Lifschitz 1986)]{landau}
{\sc Landau L. D. \& Lifschitz E. M.} {1986},
   {\it Theory of Elasticity}, 
  {Reed, Oxford, 3$^{rd}$ ed.}.

\bibitem[(Langer \& Liu 1997)]{langer}
{\sc Langer  S.A. \& Liu  A.J.} {1997},
{\it J. Phys. Chem. B} {\bf 101}, { 8667}.

\bibitem[(Larson 1997)]{larson}
{\sc Larson R.} {1997},
{\it J. Rheol.} {\bf 41}, {365}.

\bibitem[(Lauridsen, Twardos \& Dennin  2002)]{lauridsen}
{\sc Lauridsen J., Twardos M. \& Dennin M.}
{2002},
   {\it Phys. Rev. Lett.}, {\bf 89}, {098303}.

\bibitem[(Liao {\it et al.} 1997)]{liao}
{\sc Liao C.L., Chang T.P., Young D.H. \& Chang C.S.} {1997},
{\it Int. J. Solids Struct.}, {\bf 34}, {4087}.

\bibitem[(Macosko 1994)]{macosko}
{\sc Macosko C.} {1994},
{\it Rheology: principles, measurements and applications},
Wiley-VCH, New-York.

\bibitem[(Mason, Bibette, \&  Weitz 1995)]{mason}
{\sc Mason T., Bibette J., \&  Weitz D.} {1995}
{\it 
Phys. Rev. Lett. }, {\bf 75},  {2051}.
 
\bibitem[(Mecke \& Stoyan 2002)]{mecke}
{\sc Mecke K. \& Stoyan D. eds.} {2002},
{\it
Morphology of Condensed Matter - Physics and Geometry of Spatially 
Complex Systems},
   Lecture Notes in Physics 600, Springer, (Heidelberg), proceedings 
of the 2nd conference ``Spatial Statistics and Statistical Physics", 
Wuppertal (Germany), March 2001.

\bibitem[(Phan-Thien 2002)]{phan}
{\sc Phan-Thien N.} {2002},
{\it Understanding viscoelasticity}, Springer, Berlin.

\bibitem[(Porte, Berret  \& Harden 1997)]{porte}
{\sc Porte G., Berret J.F. \& Harden J.L.} {1997},
{\it  J. Phys. II France}, {\bf 7}, {459}.

\bibitem[(Poynting 1909)] {poynting}
{\sc Poynting J.H.}  {1909},
{\it Proc. Roy. Soc. London},  {\bf A82}, 546.

\bibitem[(Pratt   \& Dennin   2003)]{pratt}
{\sc   Pratt E. \& Dennin  M.}
{2003},
{\it Phys. Rev. E}, {\bf 67}, {051402}.

\bibitem[(Princen 1983)]{princen}
{\sc  Princen   H. M.} {1983},
{\em J. Coll. Interf. Sci.} {\bf 91} {160}.

\bibitem[(Reinelt   \& Kraynik  2000)]{kraynik2000}
{\sc Reinelt D. A. \& Kraynik A.}
{2000},
    {\it J. Rheology}, {\bf 44}, {453}.

\bibitem[(Rosenkilde 1967)]{rosenkilde}
{\sc Rosenkilde C.E.} {1967}, {\it J. Math. Phys.}, {\bf 8}, {84}.

\bibitem[(Rouyer {\it et al.} 2003)]{rouyer}
{\sc Rouyer F., Cohen-Addad S., Vignes-Adler M. \& H\"ohler R.}
{2003}, {\it Phys. Rev. E}, {\bf 67}, {021405}.

\bibitem[(Scheibert \& Debr\'egeas  2004)]{debregeas4}
{\sc  Scheibert J., Debr\'egeas G.} 
{2004},
in preparation.

\bibitem[(Tanguy  {\it et al.} 2002)]{tanguy}
{\sc  Tanguy A., Wittmer J.-P.,   Leonforte F.  \& Barrat J.-L.}  
{2002},
{\it Phys. Rev. B}, {\bf 66},  {174205}.


\bibitem[(Tanner \& Tanner 2003)]{hencky}
{\sc Tanner R.I. \& Tanner E.} {2003},
{\it Rheo. Acta}, {\bf 42}, {93}.

\bibitem[(Weaire \& Hutzler  1999)]{hutzler}
    {\sc Weaire D. \& Hutzler S.}
       {1999},
{\it Physics of Foams},
   {Oxford Univ. Press, Oxford}.

\bibitem[(Weaire \& Hutzler  2003)]{dilatancy}
    {\sc Weaire D. \& Hutzler S.}
       {2003},
{\it Phil. Mag.}, {\bf 83}, { 2747}.

\bibitem[(Weaire  {\it et al.}  2004)]{morgan}
{\sc Weaire D., Kern N., Cox S.J., Sullivan J.M. \& Morgan F.} {2004},
{\it Proc. Roy. Soc.: Math. Phys.}, {\bf 460}, {569}.

\bibitem[(Ybert \& di Meglio  2002)]{ybert}
{\sc  Ybert C. \& di Meglio  J.-M.}
{2002},
{\it C.R. Physique}, {\bf 3}, {555}.

\bibitem[(Zimmerman 1999)]{zimmerman}
{\sc Zimmerman J.A.} {1999},
  PhD thesis, Univ. Stanford, chap. 4, pp. 67-106, unpublished.

\end{thebibliography}
\end{document}